\documentclass[aps,amsmath,preprint, nofootinbib, superscriptaddress,]{revtex4}

\usepackage{cancel}
\usepackage{graphicx}
\usepackage{color}
\usepackage{amsmath}
\usepackage{amsfonts}

\newcommand{\beq}{\begin{eqnarray}}
\newcommand{\eeq}{\end{eqnarray}}

\newcommand{\Slash}[1]{\ooalign{\hfil/\hfil\crcr$#1$}}

\usepackage{comment}
\usepackage{epstopdf}

\def\simge{\mathrel{%
   \rlap{\raise 0.511ex \hbox{$>$}}{\lower 0.511ex \hbox{$\sim$}}}}
\def\simle{\mathrel{
   \rlap{\raise 0.511ex \hbox{$<$}}{\lower 0.511ex \hbox{$\sim$}}}}
\def\bigs{\mathrel{
   \rlap{\raise 0.531ex \hbox{$>$}}{\lower 0.531ex \hbox{$<$}}}}

\usepackage[normalem]{ulem}  

\begin{document}

\begin{flushright}
KEK-TH-1865
\end{flushright}

\title{\Large Magnetically induced QCD Kondo effect \vspace{6mm}}

\author{\large Sho Ozaki}
\email[E-mail: ]{sho@post.kek.jp}
\affiliation{Theory Center, IPNS, High Energy Accelerator Research Organization (KEK), 
1-1 Oho, Tsukuba, Ibaraki 305-0801, Japan}
\author{\large Kazunori Itakura}
\email[E-mail: ]{kazunori.itakura@kek.jp}
\affiliation{Theory Center, IPNS, High Energy Accelerator Research Organization (KEK), 
1-1 Oho, Tsukuba, Ibaraki 305-0801, Japan}
\affiliation{Graduate University for Advanced Studies (SOKENDAI),
1-1 Oho, Tsukuba, Ibaraki 305-0801, Japan}
\author{\large Yoshio Kuramoto}
\email[E-mail: ]{
yoshio.kuramoto@kek.jp}
\affiliation{Institute of Materials Structure Science, High Energy Accelerator Research Organization (KEK), 
1-1 Oho, Tsukuba, Ibaraki 305-0801, Japan\vspace{0.6cm}}

\begin{abstract}
The ``QCD Kondo effect" 
stems from the color exchange interaction in QCD with non-Abelian property, and can be realized in a high-density quark matter containing heavy-quark impurities.
We propose a novel type of the QCD Kondo effect induced by a strong magnetic field. 
In addition to the fact that the magnetic field does not affect the color degrees of freedom, two properties caused by the Landau quantization in a strong magnetic field are essential for the ``magnetically induced QCD Kondo effect"; (1)
dimensional reduction to 1+1-dimensions, and (2)
finiteness of the density of states for lowest energy quarks. We demonstrate that, in a strong magnetic field $B$, the scattering amplitude of a massless quark off a heavy quark impurity indeed shows a characteristic behavior of the Kondo effect. 
The resulting Kondo scale is estimated as $\Lambda_{\rm K} \simeq \sqrt{e_qB}\  \alpha_{s}^{1/3} {\rm{exp}}\{-{4}\pi/N_{c} \alpha_{s} {\rm{log}}( 4 \pi/\alpha_{s}) \}$ where $\alpha_{s}$ and $N_c$ are the fine structure constant of strong interaction and the number of colors in QCD, and $e_q$ is the electric charge of light quarks. 
\end{abstract}

\maketitle

\section{Introduction}

In order to understand the long-standing puzzle of enhanced resistivity in impure metals with decreasing temperature, J. Kondo  proposed in 1964 a theoretical explanation based on the third-order perturbation theory of the $s$-$d$ interaction model \cite{Kondo:1964nea}. Such anomalous behaviors of electrons induced by the presence of magnetic impurities are now called the Kondo effect. While the essential mechanism of the Kondo effect may be identified in this first theoretical analysis, a lot of theoretical investigations have been performed since then to further understand the Kondo effect.
It is now recognized that the Kondo effect provides deep and crucial insights in modern physics. Indeed,  the Kondo effect can be regarded as the first nontrivial example of renormalization groups showing the asymptotic freedom, which was found well before the discovery in QCD in the early 1970s.  
Among later developments, we mention the recognition that the orbital quantum number of localized electrons may play the role of an internal degrees of freedom \cite{Nozieres:1980}, analogous to the color of a quark.

It is well known that the Kondo effect occurs as an interplay among the following three ingredients \cite{Yamada:2004}: in addition to the existence of heavy impurities, 
(i) existence of a Fermi surface, 
(ii) quantum fluctuations (loop effect), and 
(iii) non-Abelian interaction.
In the ordinary Kondo effect in condensed matter physics, the non-Abelian interaction corresponds to a (pseudo-)spin flip interaction between a fermion near the Fermi surface and an impurity. However, other types of non-Abelian interaction are also possible. Two of the present authors together with others have recently proposed the ``QCD Kondo effect" in which non-Abelian interaction is provided by the color exchange interaction mediated by gluons \cite{Hattori:2015hka} (see also Ref.~\cite{Yasui:2013xr} for similar effects induced by contact interaction with color/isospin exchange property). In this case, the Kondo effect occurs in a high-density matter made of light quarks containing heavy quarks as impurities. 
The authors of Ref.~\cite{Hattori:2015hka} explicitly showed within perturbative renormalization group of QCD that a {\it logarithmic enhancement} indeed appears in the scattering amplitude of a light quark near the Fermi surface off the heavy quark impurity, and computed {\it the Kondo scale} (the Landau pole) where the effective coupling strength between a light quark and an impurity diverges. It is expected that the QCD Kondo effect would be relevant in medium-energy heavy ion collisions and at the core of neutron stars both of which could create high-density quark matters. Besides, since the QCD Kondo effect would affect the other many-body phenomena such as the color superconductivity, it could require modification of the QCD phase diagram in particular around the high-density region where heavy quarks start to appear. 

We can say that the QCD Kondo effect is a kind of extension of the conventional Kondo effect, and such a possibility of extending the notion of the Kondo effect urges us to further think of other possible types of the Kondo effect. For example, considering the existence of the Fermi surface, one finds what is truly essential is the {\it existence of gapless excitations} with {\it finite density of states} at the lowest energy. 
Interplay of this feature and the non-Abelian interaction gives rise to multiple particle-hole excitations near the Fermi surface and yields the logarithmic enhancement of the scattering amplitude. Therefore, if these conditions are satisfied in a system, one can expect the Kondo effect to occur. 
Notice that analogous situation can be found in other many-body phenomena such as superconductivity and magnetic catalysis  (chiral symmetry breaking induced by a magnetic field) \cite{Suganuma:1990nn, Gusynin:1994re, Gusynin:1994xp, Gusynin:1999pq, Miransky:2002rp} in which, however, non-Abelian nature of the interaction is an option. In particular,  it was explicitly pointed out in Ref.~\cite{Gusynin:1994xp} that imposing a magnetic field essentially plays the same role as the Fermi surface does in superconductivity. Thus we naturally expect that the QCD Kondo effect would be realized in the presence of a magnetic field.\footnote{Of course, this applies only when the non-Abelian interaction is not severely affected by a magnetic field, and thus does not apply to the conventional Kondo effect by the spin flip interaction.}

Let us briefly explain why imposing a strong magnetic field plays the same role as forming the Fermi surface. In a strong magnetic field, motion of a charged particle is confined to the lowest Landau level (LLL) and is effectively restricted to the direction of the magnetic field. This phenomenon is called dimensional reduction, and the fermions in a strong magnetic field are regarded as in 1+1-dimensions. If one considers massless charged fermions in a strong magnetic field, the Landau quantization of transverse motion leads to a nonzero density of states and a linear dispersion in the magnetic field direction. The claim of Ref. \cite{Gusynin:1994xp} is that the magnetic catalysis occurs by the particle-antiparticle instability at the energy surface $E=0$ of the LLL with finite density of states, which is very similar to the role played by the Fermi surface in superconductivity. Thus, one may expect the first ingredient for the Kondo effect to occur by the presence of a strong magnetic field. 

In this paper, we show that the Kondo effect induced by a strong magnetic field indeed  appears in massless QCD. 
The following three conditions are necessary: (i) a strong magnetic field, (ii) quantum fluctuations (loop effect), and (iii) color exchange interaction mediated by a gluon. 
This novel type of the Kondo effect can be referred to as the ``magnetically induced QCD Kondo effect." 
As we approach toward the Fermi energy, the scattering amplitude 
encounters the Landau pole which we call the Kondo scale. It is given by the distance from the Fermi energy as
\beq
\Lambda_{\rm K}
&\simeq&
\sqrt{e_{q}B}\,  \alpha_{s}^{\delta}\, {\rm{exp}} \left\{ - \frac{ 4 \pi }{ N_{c} \alpha_{s} {\rm{log}} \left( {4 \pi}/{\alpha_{s}} \right) } \right\} \, ,
\label{KondoScaleIntro}
\eeq
where $e_{q}$ is the electric charge of the light quark, $\alpha_{s}$
is the fine structure constant of strong interaction, and
$N_{c}$ is the number of colors.
As for the prefactor $\delta$, we get $\delta = 1/3$ in the current analysis.
But the numerical value of $\delta$ could depend on approximations we adopt.
Interestingly, this form of the Kondo scale (\ref{KondoScaleIntro}) is quite similar to the dynamical quark mass $m_{\rm{dyn}}$ induced by magnetic catalysis in QCD \cite{Miransky:2002rp}.

As we will discuss later, we assume that a light quark already acquires a dynamical mass $m_{\rm dyn}$ due to the magnetic catalysis and that the magnetic catalysis itself is not, or hardly if any, affected by the magnetically induced QCD Kondo effect. These assumptions would be fine as long as we are interested in the behavior of scattering amplitudes with decreasing energy from the high energy side and in computing the Kondo scale. This is because $m_{\rm dyn}$ is much smaller than the typical scale of the problem $\sqrt{e_qB}$ and we are allowed to treat the light quarks as massless. 
However, taking the possible ambiguities coming from approximations into account, we may expect that the similar parametric dependence of $\Lambda_{\rm K}$ and $m_{\rm dyn}$ suggests that these two scales could appear in a similar energy scale and thus leads to
a competition between the two phenomena in the energy region around these scales. 
Consider the quark-antiquark pairing enhanced by a strong magnetic field due to magnetic catalysis. If one adds heavy quark impurities to the system, then 
owing to the magnetically induced QCD Kondo effect, the heavy quark impurity attracts light quarks to inhibit the formation of chiral condensate. Thus it is expected that the magnetically induced QCD Kondo effect will weaken the magnetic catalysis. This competition possibly affects the chiral phase transition of QCD in the presence of the strong magnetic fields. 
We leave this interesting problem as a future issue.

The present paper is organized as follows:
In Sec. II, we first define the setup of massless QCD in strong magnetic fields with the quark chemical potential and a heavy quark impurity. Then we compute the scale dependence of the amplitude for the scattering between the massless quark and the heavy quark impurity in the presence of a strong magnetic field. 
In Sec. III, based on the one-loop results obtained in the previous section,
we derive the renormalization group equation for the effective coupling 
strength and solve it with an appropriate initial condition.
Here, the effective coupling is defined for the quark-impurity scattering reduced to 1+1 dimensions.
From the Landau pole of the amplitude, we estimate the Kondo scale below which the system becomes nonperturbative. We also discuss its similarities to the gaps in superconductivity and magnetic catalysis. In Sec. IV, we discuss possible applications of the magnetically induced QCD Kondo effect. Finally we summarize our study and conclude in Sec. V. 
In the Appendix, we explain how the dimensional reduction of the scattering amplitude to 1+1 dimensions occurs in strong magnetic fields.


\section{Scale dependence of the scattering amplitude in strong magnetic field}

Effective picture of low energy excitations is provided by the renormalization group (RG) equations which describe the change of interactions under the change of energy scales of interest. In particular, if the interactions are weak enough at the initial energy scales, we are able to compute the RG equations in perturbation theory. Typical examples include the color superconductivity \cite{Evans:1998ek, Hsu:1999mp, Son:1998uk} and the QCD Kondo effect \cite{Hattori:2015hka}, in both of which the smallness of the QCD coupling is guaranteed by a large chemical potential $\mu$, and the RG proceeds toward the Fermi surface (the lowest energy point). We are going to derive the RG equation for the magnetically induced QCD Kondo effect. Notice that, instead of a large chemical potential, we have now a large magnetic field $B$. As we will discuss later, the initial energy scale of the scattering problem can be taken as the scale of the order of $\sqrt{e_qB}$, and thus the QCD coupling is small enough to justify the perturbative calculation. Below, we will perturbatively compute the scattering amplitude between a light quark and a heavy quark impurity up to the one loop level and study the scale dependence of the amplitude, which are then followed by derivation of the RG equation in the next section.

\subsection{Scales in the problem}

It is instructive to summarize the relevant scales in the problem. We consider a scattering of a light quark off a heavy quark in a strong magnetic field $B$. Light quarks in a strong magnetic field undergo the magnetic catalysis and acquire a dynamical mass $m_{\rm dyn}$. It is parametrically much smaller than $\sqrt{e_qB}$, but increases with increasing $B$ \cite{Miransky:2002rp} and thus for sufficiently strong $B$, we can take $\Lambda_{\rm QCD}\ll m_{\rm dyn}\ll \sqrt{e_qB}$. Gluon propagation also gets modified in a strong magnetic field through a one loop diagram of light quarks. As a result, gluons acquire a ``screening mass" $m_g$ which is parametrically smaller than $\sqrt{e_qB}$ by a coupling constant $g$ but is larger than $m_{\rm dyn}$ (see below for more details). Therefore, in a vacuum under a strong magnetic field, we find the following hierarchy of scales: $\Lambda_{\rm QCD}\ll m_{\rm dyn}\ll m_g \ll \sqrt{e_qB}$. 
This is also consistent with the perturbative treatment with a small QCD coupling $\alpha_s\ll 1$. In the presence of a Fermi surface, which is in fact a point with the Landau quantization, we have another dimensionful parameter, a chemical potential $\mu$ or a Fermi level. We assume that $\mu \gg m_{{\rm{dyn}}}$, so that light quarks can be treated as massless above the Fermi level. 
Moreover, we take $\mu \ll \sqrt{e_qB}$ so that the gluon screening is predominantly given by the strong magnetic field.
Lastly, we have a heavy quark mass $M$, but we take the large mass limit $M\to \infty$, and $M$ does not appear in the final results.

\begin{figure*}[t]
\begin{tabular}{cc}
\begin{minipage}{0.55\hsize}
\includegraphics[width=0.7 \textwidth, bb = 160 50 750 600]{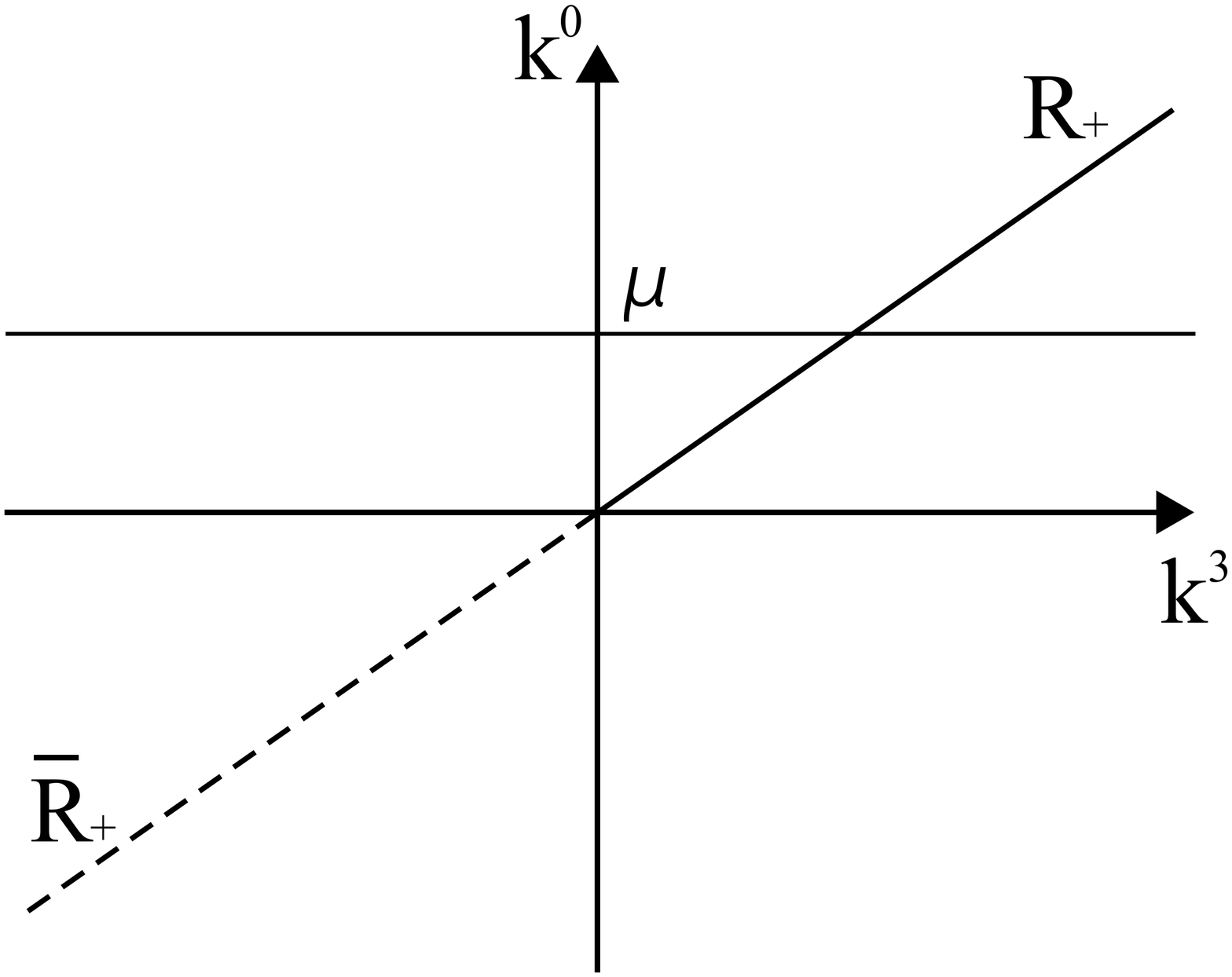}
\end{minipage}
\begin{minipage}{0.55\hsize}
\includegraphics[width=0.7 \textwidth, bb = 160 50 750 600]{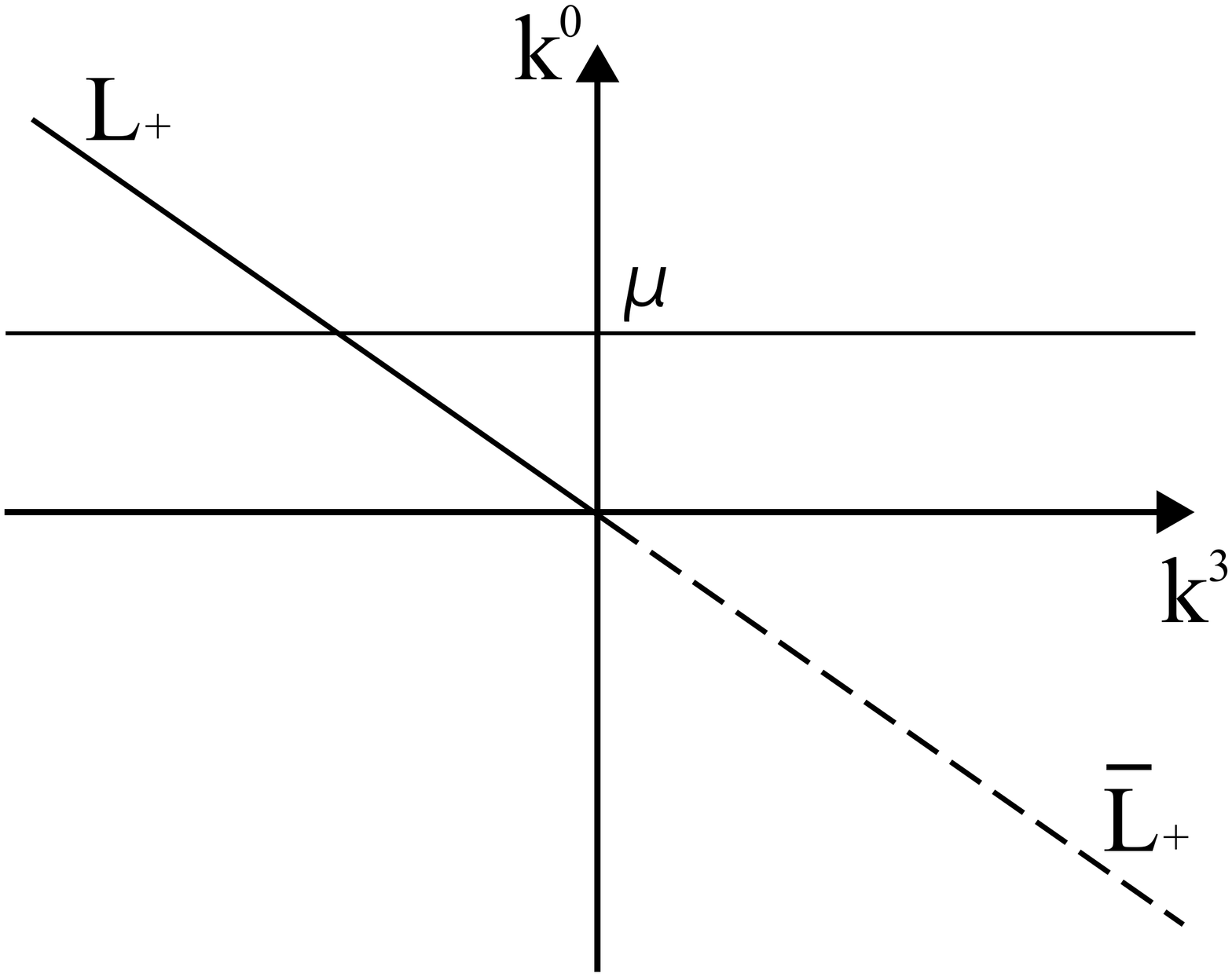}
\end{minipage}
\end{tabular}
\caption{Dispersions of massless quarks in the LLL with right and left handed chiralities are shown in the $k^0$-$k^3$ plane where the magnetic field is imposed in the third ($z$) direction. The signs $\pm$ stand for the eigenvalues of the spin in the $z$ direction. }
\label{Fig:dispersion}
\end{figure*}

Let us comment more about the chemical potential.
Although we work with a nonzero value of a chemical potential, 
it will turn out that the result is independent of the chemical potential as long as we treat the light quarks as massless. This can be easily understood from the following observation. As we mentioned in the Introduction, a massless quark (antiquark) in the LLL has a linear dispersion $k^0=\pm k^3$ as of 1+1-dimensions (see Fig.~\ref{Fig:dispersion}). For any value of the chemical potential $\mu$, the shape of dispersion nearby is always the same, and thus we can absorb the effects of a chemical potential by shifting the origin of the energy. Thus, the dominant fluctuation leading to the Kondo effect is the particle-hole excitation near the Fermi level, but the result is independent of its location.
Thus, a relevant dimensionful parameter that characterizes scattering processes is the magnetic field strength.\footnote{The heavy quark mass $M$ is taken to be infinity, and does not enter the results.}
Since the magnetic field $B$ always appears as the combination $e_qB$ with $e_q$ being the electric charge of light quarks, the QCD coupling $\alpha_s(Q^2)=g^2/4\pi $ may be evaluated at the scale given by $Q^2\simeq e_q B$.
As long as we consider a strong magnetic field $e_qB\gg \Lambda_{\rm QCD}^2$, we can take the QCD coupling small enough $\alpha_s(e_q B)\ll 1$ thanks to the asymptotic freedom, which justifies a perturbative calculation for the scattering amplitude.

\subsection{Setup}

In order to calculate the scattering amplitude, we use the QCD Lagrangian for a massless quark and a heavy quark with a mass $M$ in the presence of an external electromagnetic field and a chemical potential $\mu > 0$ (for a massless quark):
\beq
\mathcal{L}_{\rm QCD+QED}
&=& \bar{q} \left( i \Slash{D}  + \mu \gamma^{0} \right) q
+ \bar{Q} \left(  i \Slash{D} - M \right) Q - \frac{1}{4} F_{\mu \nu}^{A} F^{A \mu \nu}\, ,
\eeq
where $q$ and $Q$ are light (massless) and heavy quarks, respectively, and $M$ is the mass of the heavy quark. We consider the case of one flavor light quark and one heavy quark, but generalization to multiflavors is straightforward. The covariant derivative contains both the gluon field $A^A_\mu\, (A=1,\cdots,N_c)$ and the external electromagnetic field $a_\mu^{\rm ext}$ as $D_{\mu} = \partial_{\mu} + ig A_{\mu}^{A} T^{A} + ie_{q/Q} a_{\mu}^{\rm ext}$. Here $g$ is the gauge coupling of strong interaction, and $T^{A}$ is a generator of the SU$(N_{c})$ group. $e_q, e_Q$ are electric charges of the massless and heavy quarks, respectively. 
In this study, we only consider a constant magnetic field in the $z$-direction $\vec B=(0,0,B)$. 
In the Landau gauge, $a_{\mu}^{\rm ext}$ can be written as
\beq
a_{\mu}^{\rm ext}
&=& \left( 0,\, 0,\, B x,\, 0\right)\, .
\eeq
The field strength tensor of the gluon field is given by $F^{A}_{\mu \nu} = \partial_{\mu} A_{\nu}^{A} - \partial_{\nu} A_{\mu}^{A} + i g f^{ABC} A_{\mu}^{B} A_{\nu}^{C}$.

In this study, as we already mentioned before, we assume the following hierarchy: $\mu \ll \sqrt{|e_{q/Q}B|} \ll M$. 
As for the light quark, we take the lowest Landau level (LLL) approximation which should be justified for the massless quark, while for the heavy quark impurity we regard it as a free particle\footnote{The heavy-quark effective theory was adopted in the analysis of the QCD Kondo effect \cite{Hattori:2015hka}. However, the use of the heavy-quark effective theory is in fact not essential and a free propagator suffices in the present case.} since the coupling of the heavy quark to 
the magnetic field is suppressed by a power of $|e_{Q}B|/M^{2} \ll 1$ owing to the condition $|e_{Q}B| \ll M^{2}$.
The fermion propagator of the LLL with a finite chemical potential $\mu$ can be factorized into the longitudinal and the transverse parts as 
\beq
\mathcal{S}_{\rm{LLL}}(r,r^{\prime})
&=& S_{\parallel}(r^{}_{\parallel}, r^{\prime}_{\parallel})\,  S_{\perp}(r^{}_{\perp}, r^{\prime}_{\perp} )\, ,\label{LLL_full}
\eeq
with the longitudinal and transverse components of coordinate $r_{\parallel}^{\mu} = (t, 0, 0, z)$ and $r^{\mu}_{\perp} = (0, x, y, 0)$. 
The transverse part of the propagator is given by 
\beq
S_{\perp}(r^{}_{\perp}, r^{\prime}_{\perp} )
&=& \int \frac{dk_{y} }{ 2\pi } \frac{ 1 }{ \pi^{1/2} l_{q} } \, {\rm{e}}^{- \frac{1}{2l_{q}^{2} } \left\{ \left(x-l_{q}^{2} k_{y}\right)^{2} + \left( x^{\prime} - l_{q}^{2} k_{y} \right)^{2}  \right\} } \, {\rm{e}}^{ik_{y} (y - y^{\prime} ) }\, ,
\eeq
where $l_{q} = 1/\sqrt{e_{q}B}$ corresponds to the Larmor radius of a charged particle having the electric charge $e_{q}$ in a magnetic field $B$. 
The longitudinal part of the propagator can be written as
\beq
S_{\parallel}(r^{}_{\parallel}, r^{\prime}_{\parallel})
&=& \int \frac{ d^{2}k_{\parallel} }{ (2\pi)^{2} }\, {\rm{e}}^{-i k_{\parallel} \cdot (r^{}_{\parallel}-r^{\prime}_{\parallel}) } \, \tilde{S}_{\parallel} (k_{\parallel}; \mu)
\eeq
with ($\varepsilon>0$ is an infinitesimal constant)
\beq
\tilde{S}_{\parallel}(k_\parallel ;\mu)
&=& \frac{ i }{ 2 \epsilon_{k} } \left\{  \frac{\theta( k^{3} - k_{\rm F} )}{ k^{0} - ( \epsilon^{+}_{k} - i \varepsilon ) }
+  \frac{ \theta(k_{\rm F} - k^{3} ) \theta(k^{3}) }{ k^{0} - (\epsilon^{+}_{k} + i \varepsilon) } -  \frac{ \theta(-k^{3}) }{ k^{0} - ( \epsilon_{k}^{-} + i \varepsilon ) }
\right\} \nonumber \\
&& \times \Big( \bar{k}^{0} \gamma^{0} - k^{3} \gamma^{3} \Big) {\mathcal P}_{0}\, ,
\label{LLL_prop}
\eeq
where $k_{\parallel}^{\mu} = (k^{0},0,0,k^{3})$ with $k^{3}=k_{z}$.
We have introduced the energies $\epsilon_{k} = |k^{3}|$, $\epsilon^{\pm}_{k} = \pm \epsilon_{k} - \mu$, and the Fermi momentum $k_{\rm F} = \mu$.
${\mathcal P}_{0}$ is a spin projection operator defined by ${\mathcal P}_{0} = (1 + i \gamma^{1} \gamma^{2}) / 2$. We have also introduced a quark energy shifted by the chemical potential: $\bar{k}^{0} = k^{0} + \mu$.
Each term in the curly brace corresponds to particle ($k^3>k_{\rm F}$), hole ($0<k^3<k_{\rm F}$), and antiparticle ($k^3<0$) contributions, respectively. 
Notice that the full LLL propagator (\ref{LLL_full}) is effectively confined to a small region $|r_\perp - r_{\perp}^{\prime}|\simle l_q$ in the transverse direction. This will lead to the dimensional reduction of the scattering amplitudes to (longitudinal) 1+1 dimensions.

The gluon propagator gets modified through a quark one-loop diagram in the presence of both a magnetic field $B$ and a chemical potential $\mu$. As we alluded before, under the condition $\mu \ll \sqrt{e_qB}$, we will only consider the screening effect due to the strong magnetic field.\footnote{Namely, the Debye screening mass due to the chemical potential is much smaller than the magnetic screening mass $m_g$ which will be defined soon.} 
We employ a noncovariant gauge for the gluon field, which was adopted by Gusynin, Miransky and Shovkovy in the context of magnetic catalysis in QED \cite{Gusynin:1999pq} and QCD \cite{Miransky:2002rp}.
In this gauge, the gluon propagator in the one-loop approximation with light quarks from the LLL is given by
\beq
\mathcal{D}_{\mu \nu}^{AB} (r,r^{\prime} | e_{q}B)
&=& \int \frac{ d^{2}p_{\parallel} d^{2} p_{\perp} }{ (2\pi)^{4} }\, {\rm{e}}^{-ip_{\parallel} \cdot (r^{}_{\parallel} - r^{\prime}_{\parallel}) + ip_{\perp} \cdot (r^{}_{\perp}-r^{\prime}_{\perp}) }\, 
\tilde{\mathcal{D}}_{\mu \nu}^{AB} (p_{\parallel}, p_{\perp} |e_{q}B)
\eeq
with
\beq
\tilde{\mathcal{D}}_{\mu \nu}^{AB} (p_{\parallel}, p_{\perp}|e_{q}B)
&=& 
- i \left( \frac{ g_{\mu \nu}^{\parallel} }{ p^{2}  -   \Pi(p_{\perp}^{2}, p_{\parallel}^{2}) } + \frac{ g^{\perp}_{\mu \nu} }{ p^{2} } - \frac{ p_{\mu}^{\perp} p_{\nu}^{\perp} + p_{\mu}^{\perp} p_{\nu}^{\parallel} + p_{\mu}^{\parallel} p_{\nu}^{\perp} }{ p^{4} } \right) \delta^{AB}\, .
\label{gluon_prop}
\eeq
Here, $g_{\mu\nu}^\parallel={\rm diag}(1,0,0,-1)$, $g_{\mu \nu}^{\perp} = {\rm diag}(0,-1,-1,0)$ are parallel and transverse components of the metric, and correspondingly, $p_\mu^\parallel\equiv g_{\mu\nu}^\parallel p^\nu =(p_0,0,0,p_3)$, $p_\mu^\perp \equiv g_{\mu\nu}^\perp p^\nu =(0,p_1,p_2,0)$ are  the parallel and transverse momenta with respect to the magnetic field. The explicit form of $\Pi(p_{\perp}^{2}, p_{\parallel}^{2})$ is given by 
\beq
\Pi(p_{\perp}^{2}, p_{\parallel}^{2})
&=& + \,{\rm{exp}}\left( - \frac{p_{\perp}^{2}}{2 e_{q}B} \right)  m_{g}^{2} \, \hat{\Pi}( p_{\parallel}^{2} / m_{\rm{dyn}}^{2} )\, ,
\label{vacuum_pol}
\eeq
where we temporarily recovered the dynamical mass $m_{\rm{dyn}}$ of the light quarks for later convenience and $m_g$ works as the ``gluon mass" \cite{Miransky:2002rp}:
\beq
m_{g}^{2}
&=& \frac{ \alpha_{s} }{ \pi } e_{q} B\, .
\label{gluon_mass}
\eeq
The explicit expression for $\hat\Pi( p_{\parallel}^{2} / m_{\rm{dyn}}^{2} )$ is given in Refs.~\cite{Loskutov:1976, Dittrich:1985yb, Calucci:1993fi, Gusynin:1995nb, Fukushima:2011nu, Hattori:2012je, Hattori:2012ny}.
It should be emphasized that there is a strong screening effect in the first term of the gluon propagator (\ref{gluon_prop}) owing to the gluon mass. 
The propagator (\ref{gluon_prop}) is an analog of that employed in analyses of the Schwinger model, namely, 1+1-dimensional QED \cite{Frishman}.

Below, we will compute the scattering amplitude at the one-loop level for the purpose of deriving the RG equation. Therefore, the scattering amplitudes are evaluated at a scale $\Lambda$, and the one-loop diagrams correspond to the quantum fluctuations in the energy scales $\Lambda < E <\Lambda_0$. Here,  $\Lambda_0$ is the initial energy scale of the RG evolution and $\Lambda$ is the scale at which we want to know the effective coupling strength (effective interaction between a light quark and a heavy impurity reduced to 1+1 dimensions). 
Since these scales are measured from the Fermi level, the RG evolution with decreasing $\Lambda$ corresponds to going down toward the Fermi level.
As for the initial energy scale $\Lambda_0$, we take the scale of the order of $\sqrt{e_qB}$ because our calculation is based on the LLL approximation for the light quarks, and this approximation is valid only up to the scale of the first Landau level $\sqrt{e_qB}$ (recall the energy spectrum of a quark $\epsilon_n(k_3) =\sqrt{(k^3)^2+e_q  B n}$\, for the $n$ th Landau level). This limitation also applies to the longitudinal momentum $k^3$, which implies that the initial scale should be taken as a scale of the order of $\sqrt{e_qB}$. On the other hand, as for $\Lambda$, 
we naively expect that we are able to go down to 
the Fermi level thanks to the hierarchy $\Lambda_{\rm QCD}\ll m_{\rm dyn} \,\ll \mu \ll m_g \ll \sqrt{e_qB}$. 
However, to which scale we can actually go down depends on how large is the effective coupling. Indeed, as we will see below, the effective coupling is small enough as long as $\Lambda$ satisfies $m_{\rm{dyn}}, \mu \ll \Lambda \leq \Lambda_0$, which justifies the perturbative calculation. Then, we will find that the effective coupling grows with further decreasing $\Lambda$ and meets the Landau pole (the Kondo scale) near the Fermi level.

\begin{figure}
\begin{minipage}{0.8\hsize}
\begin{center}
\includegraphics[width=0.7 \textwidth]{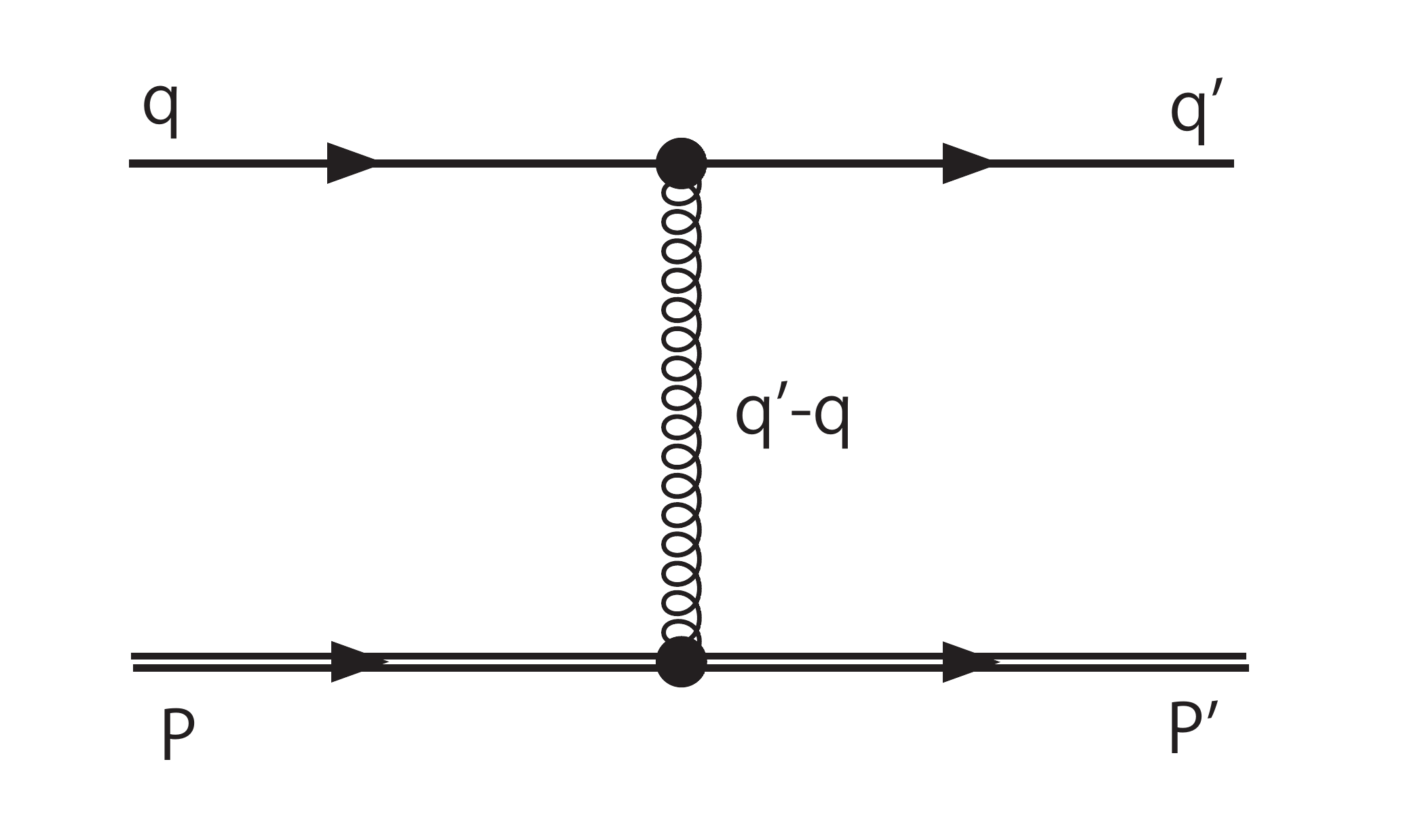}
\vskip -0.1in
\end{center}
\end{minipage}
\caption{
The tree diagram. 
Solid and double solid lines are massless and heavy quarks, respectively.
}\label{Fig:tree}
\end{figure}

\subsection{Tree amplitude and 1+1 dimensional effective coupling}

Now we compute the amplitude for scattering between a light (massless) quark near the Fermi level and a heavy quark impurity.
Under the strong magnetic field, the light quark moves only in the  direction parallel to the magnetic field. 
In the LLL with $e_{q}>0$, the spin of the light quark is fixed to the magnetic field direction.
We set the momentum of the initial quark as positive direction of the $z$-axis: $q_{z} > 0$.
Then, the leading order amplitude as shown in Fig.~\ref{Fig:tree} is given by
(see Appendix for derivation)
\beq
-i \mathcal{M}_{0}^{\rm{LLL}}
&=& (ig)^{2} \int \frac{ d^{2} Q_{\perp} }{ (2 \pi)^{2} } \left[ \bar{u}_{\rm LLL}(q^{\prime}_{\parallel}) \gamma^{\mu} (T^{A})_{a^{\prime} a} u_{\rm LLL}(q_{\parallel})\right]
\tilde{\mathcal{D}}^{AB}_{\mu \nu}(q^{\prime}_{\parallel}-q_{\parallel}, Q_{\perp} |e_{q}B)\, {\rm{e}}^{- \frac{ Q_{\perp}^{2} }{ 4 e_{q}B } } \nonumber \\
&& \times \left[\bar{U}(P^{\prime}) \gamma^{\nu} (T^{B})_{b^{\prime} b} U(P)\right]\, ,
\label{tree} 
\eeq
where the color indices of quarks can take $a, a^{\prime}, b, b^{\prime} = 1, 2, \cdots N_{c}$ and $Q_{\perp}$ is the transverse component of the gluon momentum.
The spinors are defined by 
$u_{\rm LLL}(q) = \mathcal{N}_{q}  \left( \chi_{\uparrow}, \ (\sigma_{z} { q_{z} }/{ q^{0} }) \chi_{\uparrow} \right)^{t} $ 
with $\sigma_{z} \chi_{\uparrow} = + \chi_{\uparrow}$, and 
$U(P) = \mathcal{N}_{Q} \left( \xi_{\sigma}, \  0 \right)^{t}$
with $\sigma_{z} \xi_{\pm} = \pm \xi_{\pm}$. $\mathcal{N}_{q}$ and $\mathcal{N}_{Q}$ are normalization constants. 
By using these spinors, we find 
$\bar{u}_{\rm LLL} \gamma^{\mu} u_{\rm LLL} = \bar{u}_{\rm LLL} \gamma^{\bar{\mu}} u_{\rm LLL}$ 
with $\bar{\mu} = 0, 3$, and 
$\bar{U} \gamma^{\nu} U = \bar{U} \gamma^{0} U$.
Then, in the gluon propagator (\ref{gluon_prop}), only the first term proportional to $g^{\parallel}_{00}$ contributes to the amplitude.

Notice that the tree amplitude (\ref{tree}) is defined so that the transverse momenta $q_\perp$ and $q'_\perp$ of a scattered light quark are integrated out.
Since the external momenta of the heavy quark impurity do not play any role in our study, we may regard the tree amplitude (\ref{tree}) as the one projected on the longitudinal space in 1+1 dimensions \cite{Kojo:2012js, Kojo:2013uua, Hattori:2015aki}. This should be contrasted with the QCD Kondo effect at finite densities \cite{Hattori:2015hka} and the color superconductivity \cite{Evans:1998ek, Hsu:1999mp, Son:1998uk}, where one performs the partial wave expansion of the amplitudes and project on, say, s-wave states. This dimensionally reduced amplitude becomes more evident if we introduce an effective coupling in 1+1 dimensions: 
\beq
G( q^{\prime}_{\parallel} - q_{\parallel} )\delta^{AB}\!
&\equiv& \int \frac{d^2 Q_{\perp} }{(2 \pi)^{2}} \, {\rm e}^{-{Q_{\perp}^2}/{4e_qB}} \Big[(ig)^2 i \tilde{\mathcal D}_{00}^{AB}(q'_{\parallel}-q_{\parallel}, Q_{\perp}|e_qB)\Big]\nonumber\\
&=&- 
\frac{g^{2}\delta^{AB}}{ (2 \pi)^{2} }\int d^{2} Q_{\perp}\, 
\frac{ {\rm e}^{ - Q_{\perp}^{2} / 4e_{q}B } }{ ( q^{\prime}_{\parallel} - q_{\parallel} )^{2} - Q_{\perp}^{2} - \Pi( Q_{\perp}^{2}, ( q^{\prime}_{\parallel} - q_{\parallel} )^{2} ) }\, . 
\label{effe1+1coupling_int}
\eeq
We have defined the effective coupling $G$ as a dimensionless quantity, which should be distinguished from the dimensionful coupling naturally introduced in the four-Fermi interaction. We will come back to this point later. 
The numerical value of this 1+1 dimensional effective coupling varies depending on the value of $q'_\parallel-q_\parallel$ as we explicitly demonstrate below.

Similarly to the RG in the QCD Kondo effect at finite densities \cite{Hattori:2015hka}, we consider the scattering of an off-shell light quark near the Fermi level. 
Since we are interested in the scattering amplitude at the scale $\Lambda$ which is measured from the Fermi level, 
we take the momenta of the light quark of the order of $\Lambda$ in addition to the Fermi momentum:
$q^{3} = k_{{\rm{F}}} + \mathcal{O}(\Lambda)$, as well as $q^{\prime \, 3} = k_{\rm{F}} + \mathcal{O}(\Lambda)$.
Then, we specify the kinematics for the energies and the momentum transfer as
\beq
q^{\prime \, 0}
= q^{0} = \epsilon_{\rm F}\, , \ \ \ \ \ 
q^{\prime \, 3}
- q^{3} \simeq \Lambda \, ,
\eeq
where the $\epsilon_{\rm{F}}$ stands for the Fermi energy: $\epsilon_{\rm{F}}=k_{\rm{F}}$.
The longitudinal momentum transfer $(q^{\prime}_{\parallel}-q_{\parallel})^{2}$ reads
\beq
(q^{\prime}_{\parallel}-q_{\parallel})^{2} \simeq - \Lambda^{2}.
\eeq
Since the $q^{\prime}_{\parallel}-q_{\parallel}$ dependence of the 1+1 dimensional effective coupling~(\ref{effe1+1coupling_int}) now turns into $\Lambda$ dependence,\footnote{A similar $\Lambda$ dependence of the gluon exchange interaction also appears in the context of the color superconductivity with the color magnetic interaction \cite{Hsu:1999mp, Son:1998uk}.} 
even the tree amplitude depends on the scale 
$\Lambda$.
As for the vacuum polarization (\ref{vacuum_pol}), the asymptotic form of $\hat \Pi(p_{\parallel}^{2} / m_{\rm{dyn}}^{2})$ for $|p_{\parallel}^{2}| \gg m_{\rm{dyn}}^{2}$ is given by \cite{Gusynin:1999pq, Miransky:2002rp}
\beq
\hat{\Pi}(p_{\parallel}^{2} / m_{\rm{dyn}}^{2}) \to 1.
\label{asympto_vp}
\eeq
Since $\Lambda \gg m_{\rm{dyn}}$, we use the asymptotic form (\ref{asympto_vp}) of $\hat{\Pi}(p_{\parallel}^{2} / m_{\rm{dyn}}^{2})$ in our analysis.
Then, the vacuum polarization in Eq. (\ref{effe1+1coupling_int}) becomes 
\beq
\Pi\left( 
        ( q^{\prime}_{\parallel} - q_{\parallel} )^{2} 
        \simeq - \Lambda^{2}, Q_{\perp}^{2} \right)
&=& {\rm e}^{ - \frac{ {Q_{\perp}}^{2} }{ 2e_{q}B } } m_{g}^{2} \, 
\hat{\Pi} \left( -\Lambda^{2} / m_{\rm{dyn}}^{2} \right) \nonumber \\
&\simeq& {\rm e}^{ - \frac{ {Q_{\perp}}^{2} }{ 2e_{q}B } } m_{g}^{2}\, .
\eeq
Since the denominator of the integrand in Eq.~(\ref{effe1+1coupling_int}) is now estimated as $\Lambda^2 + {\rm e}^{ -{ {Q_{\perp}}^{2} }/{ 2e_{q}B } } m_{g}^{2} {+} {Q_{\perp}}^2 $ and becomes $\sim \Lambda^2 {+} m_g^2$ in the limit ${Q_{\perp}}\to 0$, we separately evaluate the integral depending on\footnote{As an initial value for the RG equation, we will take $\Lambda=\Lambda_0$ of the order of $\sqrt{e_qB}$, but here we allow $\Lambda$ much smaller than $\sqrt{e_qB}$ for consistency check with the RG evolution. See discussion in the next section.} whether (I) $m_{g} < \Lambda$ or (II) $\Lambda < m_g$.   
Performing the integral of the transverse momentum in Eq.~(\ref{effe1+1coupling_int}), we get the  $1+1$ dimensional effective coupling as
\beq
G 
\simeq \left\{  \begin{array}{ll}
{\alpha_{s}} {\rm{log}} \left( \frac{ {4} e_{q}B }{ \Lambda^{2} } \right) & \ \ \ \ \ \ {\rm{(I)}} \\
{\alpha_{s}} {\rm{log}} \left( \frac{ {4} e_{q}B }{ m_{g}^{2} } \right) &  \ \ \ \ \ \ {\rm{(II)}}
  \end{array} \right.
 \label{Effective1+1GluonExchange}
\eeq
within the leading log accuracy. 
The effective coupling in case (I) explicitly depends on the scale $\Lambda$, and so does the tree amplitude.
On the other hand, in case (II), the effective coupling does not depend on the scale $\Lambda$, and the tree amplitude dose not give a contribution to the RG equation.

As we will immediately see below, we will encounter the same 1+1 dimensional effective coupling $G$ in the next-to-leading order and, in general, in higher orders. Thus, what truly matters in the perturbative calculation is the magnitude of $G$ rather than $\alpha_s$ or $g$ which is always small enough in our calculation. 
Notice that the $G$ can be taken small enough if one considers strong magnetic fields [so that $G \simeq \alpha_{s}\log (1/\alpha_{s})\ll 1$ in case (II) and even smaller in case (I)]. Thus, as long as we consider a strong magnetic field, the effective coupling $G$ in Eq.~(\ref{Effective1+1GluonExchange}) is small enough and the perturbative calculation is justified.

With the 1+1 dimensional effective coupling $G$, the tree amplitude (\ref{tree}) is now manifestly expressed with respect to dimensionally reduced quantities in 1+1 dimensions: 
\beq
-i  \mathcal{M}_{0}^{\rm {LLL}}
&=& -i G \left[ \bar{u}_{\rm LLL}(q^{\prime}_{\parallel}) \gamma^{0} (T^{A})_{a^{\prime} a} u_{\rm LLL}(q_{\parallel}) \right]\, \left[ \bar{U}(P^{\prime}) \gamma^{0} (T^{A})_{b^{\prime} b} U(P)\right] \nonumber \\
&=& -i G \, (T^{A})_{a^{\prime}a } (T^{A})_{b^{\prime} b}\, 
{\left[ \bar{u}_{\rm LLL}(q^{\prime}_{\parallel}) \gamma^{0}  u_{\rm LLL}(q_{\parallel}) \right] }\, 
\mathcal{N}_{Q}^{2} \xi^{\dagger}_{\sigma^{\prime}} \xi^{}_{\sigma}\, .  
\label{Leading_amp}
\eeq
In the large mass limit of the heavy quark impurity: $M \to \infty$, the heavy-quark spin is frozen as $\xi^{\dagger}_{\sigma^{\prime}} \xi^{}_{\sigma} =  \delta_{\sigma^{\prime} \sigma}$ (thus does not play a role in the QCD Kondo effect). 
This is the result of the tree amplitude and corresponds to the leading order contribution with respect to the effective coupling $G$.

\begin{figure*}[t]
\begin{tabular}{cc}
\begin{minipage}{0.6\hsize}
\includegraphics[width=0.8 \textwidth, bb = 160 50 750 600]{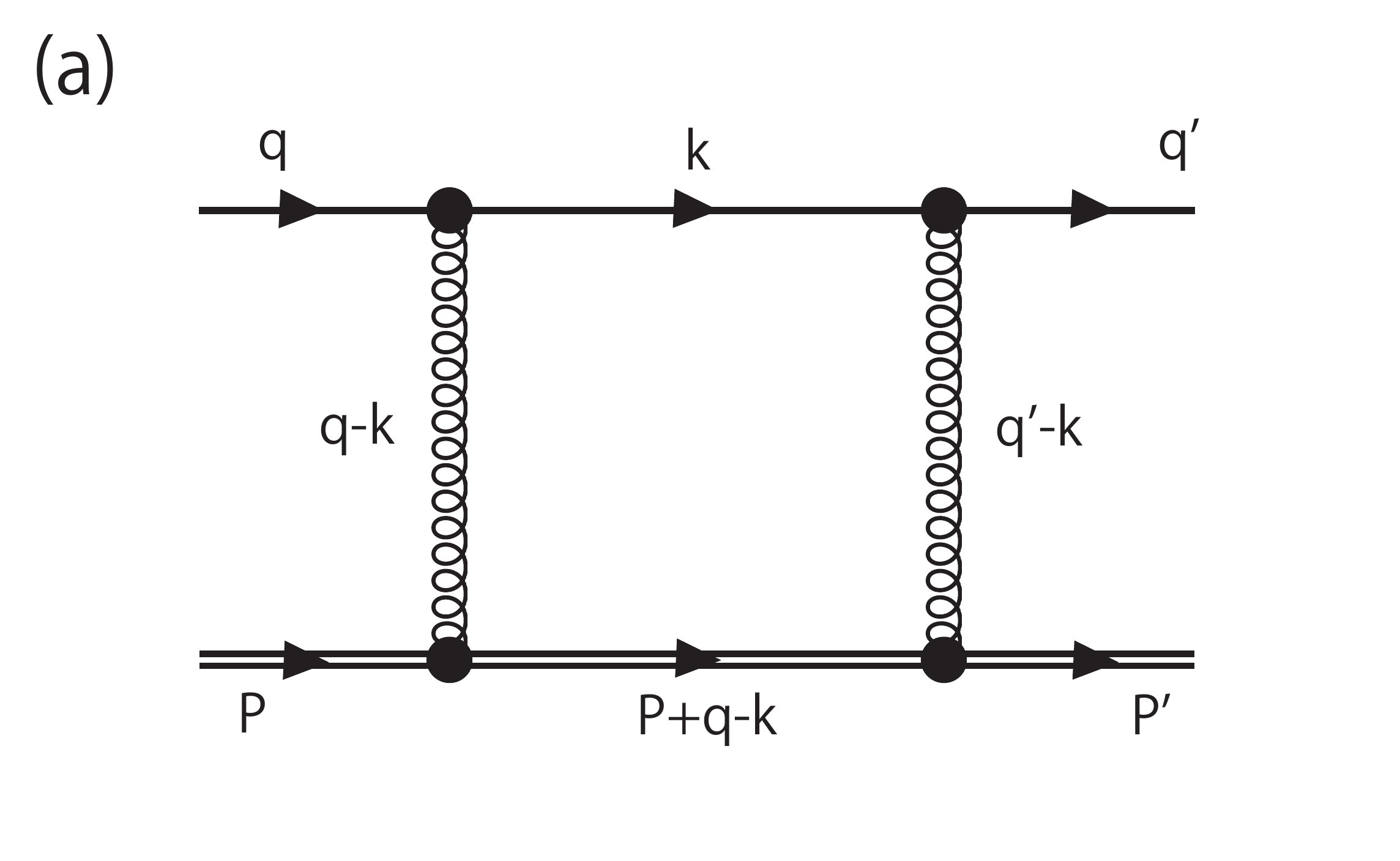}
\end{minipage}
\begin{minipage}{0.6\hsize}
\includegraphics[width=0.8 \textwidth, bb = 160 50 750 600]{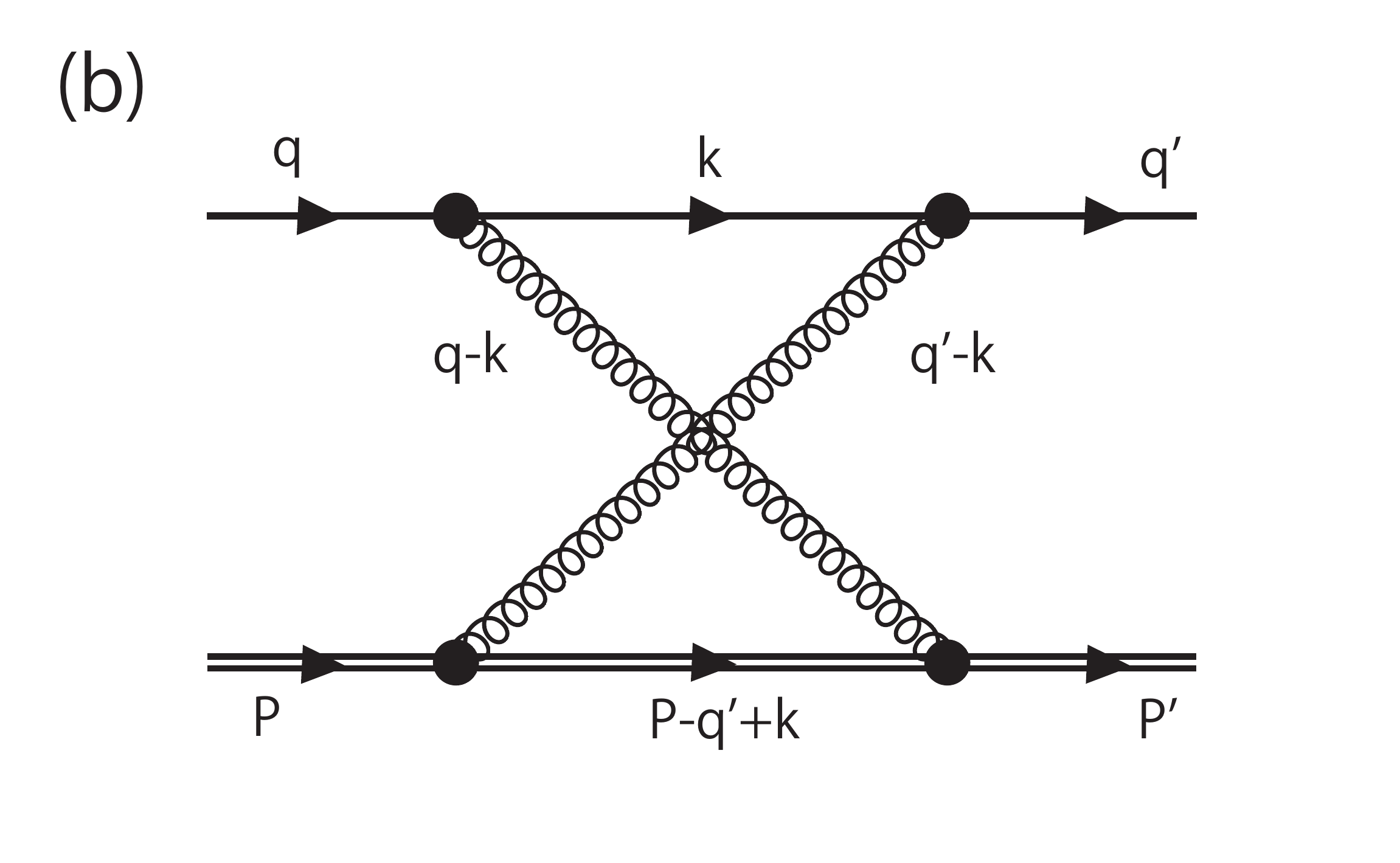}
\end{minipage}
\end{tabular}
\caption{ One-loop diagrams (a) a box diagram, (b) a crossed diagram.
}\label{Fig:one-loop}
\end{figure*}

\subsection{One-loop amplitudes}

As for the next-to-leading order, the two one-loop diagrams depicted in Figs.~\ref{Fig:one-loop}(a) and \ref{Fig:one-loop}(b) contribute to the scattering amplitude. 
Following the method similar to that for the tree amplitude, we obtain the dimensionally reduced
one-loop amplitude of the diagram (a) 
as (see the Appendix for details)
\beq
&&\hspace{-1cm}-i \mathcal{M}^{(a) {\rm LLL}}_{\rm 1-loop}\nonumber\\
&=& (ig)^{4} \int \frac{d^{{2}}k_{{\parallel}}}{(2\pi)^{{2}} }\,  
\Big[ \bar{u}_{\rm LLL}(q^{\prime}_{{\parallel}}) \gamma^{\mu} (T^{A})_{a^{\prime}a^{\prime \prime}}  {\tilde{S}_{\parallel} (k_{\parallel};\mu) } \gamma^{\nu} (T^{B})_{a^{\prime \prime} a} u_{\rm LLL}(q_{{\parallel}})\Big]\nonumber \\
&&\hspace{-3mm} \times \ {\int \frac{ d^{2} Q_{\perp}^{\prime} }{ (2\pi)^{2} }} \tilde{\mathcal{D}}_{\mu \sigma }^{AC} ( q^{\prime}_{{\parallel}} - \bar{k}_{{\parallel}}, {Q_{\perp}^{\prime} } | e_{q}B) \, 
{{\rm{e}}^{ - \frac{ Q_{\perp}^{\prime 2} }{ 4 e_{q}B } } }
{\int \frac{ d^{2} Q_{\perp} }{ (2\pi)^{2} }} \tilde{\mathcal{D}}_{\nu\rho}^{BD} (q_{{\parallel}}-\bar{k}_{{\parallel}}, {Q_{\perp}} |e_{q}B)
\, {{\rm{e}}^{ - \frac{ Q_{\perp}^{2} }{ 4 e_{q}B } } } 
\, {{\rm{e}}^{ - i\frac{ \vec{Q}_{\perp} \times \vec{Q}_{\perp}^{\prime} }{  2e_{q}B } } } \nonumber \\
&&\hspace{-3mm}  \times \Big[\bar{U}(P^{\prime}) \gamma^{\sigma} (T^{C})_{b^{\prime} b^{\prime \prime}} {\tilde{S}_{\rm{H}}}(P+q-\bar{k}) \gamma^{\rho} (T^{D})_{b^{\prime \prime} b} U(P)\Big]\, ,
\label{one-loop-amp-a} 
\eeq 
where $\bar{k}^{\mu} = (\bar k^{0}, \vec{k})=(k^{0} + \mu, \vec{k})$, and $\tilde{S}_{\rm H}$ is the heavy quark propagator in the momentum space.
The two square brackets (in the first and third lines) correspond to light and heavy quark lines, respectively, and the second line corresponds to gluons exchanged between light and heavy quarks. Again, (apart from the heavy quark part) the amplitude depends only on longitudinal momenta $q_\parallel$ and $q'_\parallel$ of the scattered light quark. The integration with respect to $k_\parallel$ corresponds to the loop integral, while the integrations over two transverse momenta $Q_\perp$ and $Q'_\perp$ should be associated with the effective coupling which appeared in the tree amplitude [see Eq.~(\ref{effe1+1coupling_int})]. However, we notice the presence of a phase factor ${\rm{exp}}{\left(-i \, { \vec{Q}_{\perp} \times \vec{Q}_{\perp}^{\prime} }/{  2e_{q}B }\right) }$ which involves $Q_\perp$ and $Q'_\perp$ and thus makes the two integrations inseparable. This magnetic phase factor is gauge invariant, which appears also in the symmetric gauge.
In fact, this phase factor can be neglected in the present approximation,\footnote{Roughly speaking, the integration with respect to the transverse momentum $Q_\perp$ can be approximated as $\int d^2Q_\perp/Q_\perp^2$ with lower and upper cutoffs given by $m_g$ (due to the denominator of the gluon propagator) and $\sqrt{e_qB}$ (due to the Gaussian factor). Since the phase factor becomes sizable when $|Q_{\perp}|, |Q_{\perp}^{\prime}|$ approach to $\sqrt{e_{q}B}$, it will modify the integration only in the region around the upper cutoff. The relative importance near the cut-off region compared to the whole integral indeed becomes negligible if the ratio $m_{g}^{2} / e_{q}B \sim \alpha_{s}$ is small enough, which has been numerically checked.} which enables us to identify the integrations over two transverse momenta in the second line of Eq.~(\ref{one-loop-amp-a}) with the effective couplings.
The light quark propagator is proportional to the spin projection operator ${\mathcal P}_{0}$, and then
we can use $\bar{u}_{\rm LLL} \gamma_{\mu} {\mathcal P}_{0} \gamma_{\nu} u_{\rm LLL} = \bar{u}_{\rm LLL} \gamma_{\bar{\mu}} {\mathcal P}_{0} \gamma_{\bar{\nu}} u_{\rm LLL}$.
In the large mass limit: $M \to \infty$, the heavy quark propagator reads 
\beq
\tilde{S}_{\rm{H}} (P+q-\bar{k}) = \frac{i}{ \Slash{P} + \Slash{q} - \bar{\Slash{k}} -M } \sim \frac{i}{q^{\prime 0} - \bar{k}^{0} } {\mathcal P}_{+}, 
\eeq
with ${\mathcal P}_{+} = (1 +  \gamma^{0})/2$ being the projection operator to the positive energy spinor.
Thus, the vertex structure on the heavy quark side becomes $\bar{U}(P^{\prime}) \gamma_{\sigma} {\mathcal P}_{+} \gamma_{\rho} U(P)
= \bar{U}(P^{\prime}) \gamma_{0} {\mathcal P}_{+} \gamma_{0} U(P)$.
Again, only the first term of the gluon propagator (\ref{gluon_prop}), proportional to $g_{0 0}$, contributes to the amplitude (\ref{one-loop-amp-a}).

Within the approximation of neglecting the magnetic phase factor, we can express the second line of Eq.~(\ref{one-loop-amp-a}) in terms of the 1+1 dimensional effective coupling $G$. This brings us to the following representation of the one-loop amplitude of the diagram (a):
\beq
-i {\mathcal{M}^{({\rm{a}}) {\rm LLL}}_{\rm 1-loop}}
&=& i^{4} \mathcal{T}_{a^{\prime} a; b^{\prime} b}^{({\rm{a}})}
\int \frac{ d^{{2}} k_{{\parallel}} }{ (2\pi)^{{2}} } 
\left[\bar{u}_{\rm LLL}(q^{\prime}_{\parallel}) \gamma_{0}\, \left( \bar{k}^{0} \gamma^{0} - k^{3} \gamma^{3} \right)
  {\mathcal P}_{0} \gamma_{0} u_{\rm LLL}(q_{\parallel})\right]
 \nonumber \\
&&\quad \times \frac{ i }{ {2}\epsilon_{k} }\left\{  \frac{ \theta (k^{3} - k_{\rm F}) }{ k^{0} - ( \epsilon^{+}_{k} - i \varepsilon)} + \frac{ \theta ( k_{\rm F} - k^{3} ) \theta(k^{3})  }{ k^{0} - ( \epsilon^{+}_{k} + i \varepsilon ) } -  \frac{ \theta(-k^{3}) }{ k^{0} - ( \epsilon^{-}_{k} + i \varepsilon) } 
\right\} \nonumber \\
&&\quad \times 
 (i G)^{2} 
\frac{ i } { q^{\prime 0} - \bar{k}^{0} }\left[\bar{U}(P^{\prime}) \gamma^{0}  {\mathcal P}_{+} \gamma^{0} U(P)\right]\, ,\  
\eeq
where the color factor $\mathcal{T}_{a^{\prime} a; b^{\prime} b}^{({\rm{a}})}$ has been defined as $\mathcal{T}^{({\rm{a}})}_{a^{\prime} a; b^{\prime} b} = (T^{A})_{a^{\prime} a^{\prime \prime} } (T^{B})_{a^{\prime \prime} a} (T^{A})_{b^{\prime} b^{\prime \prime}} (T^{B})_{b^{\prime \prime} b }$.
The second line comes from the light quark propagator $\tilde S_\parallel (k_\parallel ; \mu)$ in the loop.
We perform the $k^{0}$ integral with the contour closed in the upper half of the complex $k^0$ plane. 
Then, picking up the contribution corresponding to particle excitation of the light quark (the first term of the light quark propagator),  we find 
\beq
-i {\mathcal{M}^{({\rm{a}}){\rm LLL}}_{\rm 1-loop}}
&\simeq & i G^{2} \mathcal{T}_{a^{\prime} a; b^{\prime} b}^{({\rm{a}})} \ 
\Big[\bar{u}_{\rm LLL}(q^{\prime}_{\parallel}) \frac{ \gamma^{0} + \gamma^{3}}{{2}} {\mathcal P}_{0}  u_{\rm LLL}(q_{\parallel})\Big] \Big[ \bar{U}(P^{\prime}) \gamma^{0} {\mathcal P}_{+} \gamma^{0} U(P)\Big] \nonumber \\
&& 
\times 
\int_{k_{\rm F} + \Lambda}^{k_{\rm F}+\Lambda_{0}} \frac{ dk^{3} }{ 2\pi } \frac{ 1 }{ \epsilon_{k} - \epsilon_{\rm F} }\ 
\, , 
\label{one-loop-amp-a2}
\eeq
where the $k^3$ integral is limited to the higher energy strip $k_{\rm F}+\Lambda \le k^3 \le k_{\rm F}+\Lambda_0$ with $k^{}_{\rm F}$ being the Fermi momentum of the light quark $k^{}_{\rm F}=\mu$.
By changing the integral variable as $k^{3} \to  K = k^{3} - k_{\rm F}$, 
we finally arrive at the expression independent of 
$k^{}_{\rm F}=\mu$: 
\beq
-i {\mathcal{M}^{({\rm{a}}){\rm LLL}}_{\rm 1-loop}}
&\simeq& i G^{2} \, \mathcal{T}_{a^{\prime} a; b^{\prime} b}^{({\rm{a}})} \ 
\Big[\bar{u}_{\rm LLL}(q^{\prime}_{\parallel}) \frac{  \gamma^{0} + \gamma^{3}  }{2} {\mathcal P}_{0}  u_{\rm LLL}(q_{\parallel})\Big] \Big[\bar{U}(P^{\prime}) {\mathcal P}_{+}  U(P)\Big]\nonumber \\
&& \times  \ \int_{{\Lambda}}^{\Lambda_0} \frac{ d{K} }{2\pi }\, \frac{ 1 }{ {K} }\, .
\label{one-loop_amp_a}
\eeq
Now we can clearly see that this amplitude has a logarithmic contribution and does not depend on the chemical potential. As we already discussed before, with the linear dispersion of the massless quarks we can always absorb the information of the Fermi level (the chemical potential) into the integral variable as we have just done so.

Next we consider the crossed diagram (b). 
We basically follow the same procedure as in the diagram (a), but with attention on the color index and momentum flow. The one-loop amplitude of the diagram (b), reduced to 1+1 dimensions, can be obtained as
\beq
-i {\mathcal{M}^{({\rm{b}}){\rm LLL}}_{\rm 1-loop}}
&=& i^{4} \int \frac{ d^{{2}}k_{\parallel} }{ (2\pi)^{{2}} } \, 
\Big[\bar{u}_{\rm LLL}(q^{\prime}_{\parallel}) \gamma^{0} (T^{A})_{a^{\prime} a^{\prime \prime} } {\tilde{S}_{\parallel}(k_{\parallel} ;\mu)} \gamma^{0} (T^{B})_{a^{\prime \prime} a} u_{\rm LLL}(q_{\parallel})\Big] \nonumber \\
&&\quad\quad\ \quad \times \ {(iG)^{2}} \Big[ \bar{U}(P^{\prime}) \gamma^{0} (T^{D})_{b^{\prime} b^{\prime \prime}} {\tilde{S}_{\rm{H}}}(P-q^{\prime} + \bar{k}) \gamma^{0} (T^{C})_{b^{\prime \prime} b} U(P)\Big]\, .
\label{one-loop_amp-b}
\eeq
Here we have already used the facts that $\bar{u}_{\rm LLL} \gamma_{\mu} {\mathcal P}_{0} \gamma_{\nu} u_{\rm LLL} = \bar{u}_{\rm LLL} \gamma_{\bar{\mu}} {\mathcal P}_{0} \gamma_{\bar{\nu}} u_{\rm LLL}$ and $\bar{U}(P^{\prime}) \gamma_{\sigma} {\mathcal P}_{+} \gamma_{\rho} U(P)
= \bar{U}(P^{\prime}) \gamma_{0} {\mathcal P}_{+} \gamma_{0} U(P)$ as in the previous calculation for the diagram (a).
We have also neglected the magnetic phase factor for the same reason as discussed below Eq.~(\ref{one-loop-amp-a}).
Performing the $k^{0}$ integral with lower contour, 
we find
\beq
- i {\mathcal{M}^{({\rm{b}}){\rm LLL}}_{\rm 1-loop}}
&\simeq& i G^{2}\, \mathcal{T}^{({\rm{b}})}_{a^{\prime} a; b^{\prime} b} \, 
\Big[ \bar{u}_{\rm LLL}(q^{\prime}_{\parallel})  \frac{ \gamma^{0} + \gamma^{3} }{2} {\mathcal P}_{0} u_{\rm LLL}(q_{\parallel})\Big] \Big[\bar{U}(P^{\prime}){\mathcal P}_{+} U(P)\Big] \nonumber \\
&& \times \ 
{\int^{k_{\rm F}-\Lambda}_{k_{\rm F}-\Lambda_0} \frac{ dk^{3} }{ 2\pi } 
\frac{1}{ \epsilon_{k} - \epsilon_{\rm F} } }, 
\label{anti-particle_cont}
\eeq
where we have defined the color factor as $\mathcal{T}^{(b)}_{a^{\prime} a ; b^{\prime} b} = (T^{A})_{a^{\prime} a^{\prime \prime} } (T^{B})_{a^{\prime \prime} a} (T^{B})_{b^{\prime} b^{\prime \prime}} (T^{A})_{b^{\prime \prime} b }$. 
As is evident from the integration range, this amplitude comes from the contribution below the Fermi momentum, $k^3<k_{\rm F}$, which in general contains both holes and antiparticles. In the present case with $\Lambda_0>\Lambda \gg k_{\rm F}=\mu$, since both $k_{\rm{F}} -\Lambda_{0}$ and $k_{\rm{F}} -\Lambda$ are negative, the integration region is in the Dirac sea and thus the amplitude comes from the antiquark contribution.\footnote{This is contrasted with the ordinary Kondo effect in condensed matter with a contact interaction, where the crossed diagram corresponds to the hole contribution.} Furthermore, thanks to the condition $\Lambda \gg m_{\rm dyn}$, the integration region is far away from the region $|k^3|<m_{\rm dyn}$, and thus we can safely treat the antiquark as massless. 
We shall change the integral variable as $k^{3} \to K =
k_{\rm F} - k^{3}$. 
Then, the amplitude becomes
\beq
- i {\mathcal{M}^{({\rm{b}}){\rm LLL}}_{\rm 1-loop}}
&\simeq&  i G^{2} \mathcal{T}^{({\rm{b}})}_{a^{\prime} a; b^{\prime} b}  
\Big[\bar{u}_{\rm LLL}(q^{\prime}_{\parallel})  \frac{ \gamma^{0} + \gamma^{3} }{2} {\mathcal P}_{0} u_{\rm LLL}(q_{\parallel}) \Big] \Big[ \bar{U}(P^{\prime}) {\mathcal P}_{+}  U(P)\Big] \nonumber \\
&& \times \, 
 \int_{{\Lambda}}^{\Lambda_0} \frac{ d{K} }{ 2\pi } \frac{1}{-{K}}\, .
\label{one-loop-amplitude_b}
\eeq
The chemical potential is again absorbed into the integral variable thanks to the linear dispersion of the massless quark.

When combining the contributions from the diagrams (a) and (b), we use the following identities for the color factors: 
\beq
\mathcal{T}^{({\rm{a}})}_{a^{\prime} a; b^{\prime} b}
&=&   \frac{ N_{c}^{2} - 1 }{ 4N_{c}^{2} } \delta_{a^{\prime} a} \delta_{b^{\prime} b} - \frac{1}{N_{c}} (T^{A})_{a^{\prime} a} (T^{A})_{b^{\prime} b}\, \label{Ta},\\ 
\mathcal{T}^{({\rm{b}})}_{a^{\prime} a; b^{\prime} b} 
&=&  \frac{ N_{c}^{2} -1 }{ 4 N_{c}^{2} } \delta_{a^{\prime} a } \delta_{b^{\prime} b}+\left(-\frac{1}{N_c}+\frac{N_c}{2}\right) (T^{A})_{a^{\prime} a} (T^{A})_{b^{\prime} b} \, .\label{Tb}
\eeq
Then, we find the total amplitude at the one-loop level:
\beq
-i {\mathcal{M}^{\rm LLL}_{\rm 1-loop}}
&=& - i \left( {\mathcal{M}^{({\rm{a}}){\rm LLL}}_{\rm 1-loop}} + {\mathcal{M}^{({\rm{b}}){\rm LLL}}_{\rm 1-loop}} \right) \nonumber \\
&\simeq& -i G^{2} \frac{N_{c}}{2}(T^{A})_{a^{\prime}a}(T^{A})_{b^{\prime} b}\, 
{\left[ \bar{u}_{\rm LLL}(q^{\prime}_{\parallel}) \gamma^{0}  u_{\rm LLL}(q_{\parallel}) \right] }
\ \mathcal{N}_{Q}^{2} \xi^{\dagger}_{\sigma^{\prime}} \xi^{}_{\sigma}\nonumber \\
&& \times   \  \int^{{\Lambda_{0}}}_{{\Lambda}} \frac{ d{K} }{  2\pi } \frac{1}{{K}}\, .
\label{total_amp1}
\eeq
We see that the logarithmic terms coming from box (\ref{one-loop_amp_a}) and crossed (\ref{one-loop-amplitude_b}) diagrams do not cancel each other, and this term remains in the total amplitude thanks to the non-Abelian property of the interaction as in the ordinary Kondo effect. In fact, this incomplete cancellation due to the difference in Eqs.~(\ref{Ta}) and (\ref{Tb}) is exactly the same as what we encounter in the QCD Kondo effect at finite density \cite{Hattori:2015hka}. While the strength of the effective interaction changes in the presence of magnetic fields, its non-Abelian structure does not, and thus works in the same way as in the QCD Kondo effect. 

Furthermore, since light quarks can be regarded as massless above the Fermi level, the helicity conservation must be satisfied. Thus, only forward scattering of the light quark near the Fermi level off the heavy quark impurity is allowed in the 1+1 dimensions.


\section{Renormalization group equation and the Kondo scale}

\subsection{Renormalization group equation}
In this section, we derive the RG equation of the effective coupling strength for the interaction between a light quark and a heavy quark.
Combining the tree amplitude (\ref{Leading_amp}) and the one-loop amplitude 
(\ref{total_amp1}), we find that the one-loop contribution gives rise to an additional logarithmic term to the tree one: 
\beq
-i{{\mathcal M}^{\rm LLL}}
&= & -i \Big( {{\mathcal M}_0^{\rm LLL}} + {{\mathcal M}_{\rm 1-loop}^{\rm LLL}} \Big) \nonumber\\
 &\simeq & -i{{\mathcal M}_0^{\rm LLL}} \left( 1 + \frac{G}{2\pi} \frac{N_c}{2} \log \frac{\Lambda_{0}}{\Lambda}  \right)\, .\label{log}
\eeq
From this result, we notice that the one-loop amplitude ${\mathcal M}_{\rm 1-loop}^{\rm LLL}$ is proportional to the tree amplitude $\mathcal{M}_0^{\rm LLL}$, and thus we can redefine the coupling $G$ with the logarithmic term included. 
From the viewpoint of the RG, this implies that the RG flow of the amplitude can be represented as that of the ``effective coupling strength" $G(\Lambda)$ at the scale $\Lambda$, and the coupling $G$ defined at the tree amplitude (\ref{Effective1+1GluonExchange}) should be treated as the initial value at $\Lambda=\Lambda_0$ \cite{Hattori:2015hka}. 
More precisely, for an infinitesimal change of the scale, the new effective coupling 
$G(\Lambda -\delta\Lambda)$ 
defined by the left-hand side of Eq.~(\ref{log}) can be expressed as 
$G(\Lambda -\delta\Lambda)\simeq G(\Lambda)-{\delta G(\Lambda')}/{\delta \Lambda'}|_{\Lambda'=\Lambda}\delta \Lambda$ 
with the first term defined by the tree amplitude and the second by the one-loop amplitude. However, we have to be careful when the tree amplitude itself has an explicit scale dependence as we have seen in Eq.~(\ref{Effective1+1GluonExchange}). In such a case, we need to include the change of the tree amplitude into the second term together with the one-loop contribution.\footnote{Otherwise the coupling does not vary in the absence of the one-loop contribution.} This happens only for case (I) $m_g<\Lambda$. Since we start from the initial scale $\Lambda_0$ of the order of $\sqrt{e_qB}$ which is much larger than $m_g$ [corresponding to case (I)] and we decrease the scale $\Lambda$ according to the RG equation, we must distinguish two regimes for the running scale $\Lambda$ corresponding to cases (I) and (II) in Eq.~(\ref{Effective1+1GluonExchange}): 
\begin{eqnarray}
&{\rm (I)}&  m_g<\Lambda \simle \sqrt{e_qB}\, ,\nonumber\\
&{\rm (II)}& m_{\rm dyn} \ll \Lambda <m_g.\nonumber
\end{eqnarray}
The lower limit for regime (II) is put because the present calculation with massless approximation is valid only for $\Lambda \gg \ m_{\rm dyn}$.
 According to Eqs.~(\ref{Effective1+1GluonExchange}) and (\ref{log}), we find the following RG equations for each regime: 
\beq
&{\rm (I)}&\quad \Lambda \frac{d}{d\Lambda}G(\Lambda)=-{2\alpha_{s} } -\frac{N_c}{4\pi}  G^2(\Lambda)\, , \label{RGeq-I}\\
&{\rm (II)}&\quad\Lambda \frac{d}{d\Lambda}G(\Lambda)=-\frac{N_c}{4\pi} G^2(\Lambda)\, ,\label{RGeq-II}
\eeq
where the first term in Eq.~(\ref{RGeq-I}) comes from the scale dependence of the tree amplitude.\footnote{A RG equation similar to Eq.~(\ref{RGeq-I}) describes color superconductivity induced by the color magnetic interaction \cite{Son:1998uk}.} As we start from the initial scale $\Lambda_0$ and decrease the scale $\Lambda$, the effective coupling $G(\Lambda)$ first obeys Eq.~(\ref{RGeq-I}) in regime (I) then Eq.~(\ref{RGeq-II}) in regime (II).

\subsection{Solutions to the RG equations and matching}

Before considering the effect of the additional contribution from the tree amplitude in regime (I), let us solve the simpler RG equation (\ref{RGeq-II}) in regime (II). Notice that it is the famous form of the differential equation yielding the asymptotic freedom as seen in QCD. We solve this equation with the initial condition $G(\Lambda_0)= {\alpha_{s}} \, {\rm{log}} ({4} e_{q}B / m_{g}^{2})\equiv G_0$ specified at $\Lambda=\Lambda_0 \simle m_g$. 
Then, the solution 
\beq
G(\Lambda)=\frac{G_0}{1+\frac{1}{4\pi}N_c  G_0
\log \left({\Lambda}/{\Lambda_0}\right)} \label{RGsolution}
\eeq
shows that the effective coupling $G(\Lambda)$ decreases with increasing $\Lambda$ (asymptotic freedom). Conversely, the effective coupling increases with decreasing $\Lambda$, and diverges at some scale. This is the Landau pole called the {\it Kondo scale} (an analog of $\Lambda_{\rm QCD}$ in QCD), and it is given by 
\beq
\Lambda_{\rm K}
&\simeq & \Lambda_{0} \ {\rm{exp}}\left(- \frac{4\pi}{ N_{c}  G_0  } \right) \label{Kondo_scale}
\eeq
with $\Lambda_{0}\simle m_{g}$. The effective coupling in the exponent is evaluated at this scale $G_0=G(\Lambda_0)$. 
Below and around the Kondo scale, the system becomes nonperturbative even if the gauge coupling $g$ is small enough. 
By using the Kondo scale (\ref{Kondo_scale}), the effective coupling can be expressed as 
\beq
G(\Lambda)=\frac{1}{\frac{1}{4\pi} N_c  \log {\Lambda}/{\Lambda_{\rm K}}}\, .
\eeq
This is again similar to the asymptotic freedom in QCD.

A few comments are in order about the Kondo scale (\ref{Kondo_scale}). First of all, let us consider the QCD Kondo effect at high density with the contact interaction whose strength is given by a dimensionful coupling $\bar G_{\rm c}$ so that the interaction is given by $\bar G_{\rm c} (\bar q \Gamma q) (\bar Q \Gamma Q).$ With the density of state $\rho^{}_{\rm{F}}$ at the Fermi surface, one obtains a Kondo scale $\Lambda_{\rm K}\propto \exp \left(-2/N_c \rho^{}_{\rm{F}}  \bar G_{\rm c}\right)$ \cite{Yasui:2013xr,Hattori:2015hka}. 
Notice that a dimensionless combination $\rho_{\rm{F}} \bar G_{\rm{c}}$ corresponds to the dimensionless coupling $G_{0} / (2\pi)$ in Eq.~(\ref{Kondo_scale}). This observation suggests that the density of state of the LLL, $\rho_{\rm LLL},$ was hidden in our definition of the effective coupling. Indeed, integration over the transverse momentum of the light quark yields $\rho_{\rm LLL}\equiv \frac{1}{2\pi}\cdot  \int \frac{d k_y}{2\pi} \frac{1}{\pi^{1/2} l_{q}} {\rm e}^{- l_{q}^{2} k_y^2} =e_qB/(2\pi)^2$.
The similarity between these two Kondo scales is a consequence of the fact that imposing a magnetic field plays the same role as having the Fermi surface. This should be compared with another similarity between the superconducting gap $\Delta \propto 
{\rm{exp}}( - c / \rho^{}_{\rm{F}}  \bar G_{\rm{s}} )$  and the chiral condensate (dynamical mass) from magnetic catalysis: $m_{\rm{dyn}} 
\propto {\rm{exp}}( -c^{\prime}/ \rho^{}_{\rm{LLL}} 
\bar G_{\rm m} )$ where 
 $\bar G_{\rm s}, \bar G_{\rm m}$ are the coupling strength of the contact interactions, and $c,c'$ are some numerical constants. The latter similarity was the crucial observation made in Ref.~\cite{Gusynin:1994xp}.  
Substituting $G_{0} = \alpha_{s} {\rm{log}} (4 e_{q} B / m_{g}^{2} ) = \alpha_{s} {\rm{log}} (4\pi / \alpha_{s})$ into Eq. (\ref{Kondo_scale}), one finds
\beq
\Lambda_{\rm K}
&\simeq & {\sqrt{e_qB}\, \alpha_{s}^{\delta} \, {\rm{exp}} \left\{ - \frac{ {4} \pi }{N_{c} \alpha_{s} {\rm{log}}({4}\pi/\alpha_{s}) } \right\}\, , }
\label{Kscale}
\eeq
where we have taken $\Lambda_0=m_g$ and thus $\delta = 1/2$.

Now we go back to the original problem with the initial scale $\Lambda_0$ taken in regime (I) and, correspondingly, the RG equation (\ref{RGeq-I}). The solution to the RG equation (\ref{RGeq-I}) is given by
\beq
G(\Lambda)
&=&  \sqrt{ \frac{ 8\pi \alpha_{s} }{ N_{c} } }  \,  
{\rm{tan}} \left[ 
{\rm{arctan}} \left( { \sqrt{ \frac{ N_{c} }{ 8\pi \alpha_{s} } } } G(\Lambda_{0}) \right) - { \sqrt{ \frac{ N_{c} \alpha_{s} }{ 8\pi } } } {\rm{log}} \frac{ \Lambda^{2} }{ \Lambda_{0}^{2} } \right] 
\eeq
with the initial effective coupling
\beq
G(\Lambda_{0})
&=& { \alpha_{s} } {\rm{log}} \left( \frac{ {4} e_{q} B }{ \Lambda_{0}^{2} } \right).
\eeq
With decreasing $\Lambda$, the effective coupling grows as expected. 
This solution is valid in regime (I) and its value at the lower limit 
 $\Lambda = m_{g}$ is evaluated as
\beq
G(m_{g})
&=& {\alpha_{s} }{\rm{log}} \frac{ {4} e_{q} B }{ m_{g}^{2}}
\left\{ 
1 + \frac{1}{3}\cdot \left(\sqrt{\frac{ N_{c} \alpha_{s} }{ 8\pi }}  \, {\rm{log}} \frac{ {4} e_{q} B }{ m_{g}^{2} }\right)^2 + \cdots 
\right\}.
\label{Gmg}
\eeq
Here we have expanded the solution with respect to $\sqrt{\alpha_s} \, {\rm{log}}( 4 e_{q}B / m_{g}^{2}) = \sqrt{\alpha_s} \, {\rm{log}}( 4 \pi / \alpha_{s} )\ll 1$ which naturally appears in the argument of tangent or arctangent, and the contribution depending on $\Lambda_0$ appears in the next subleading term.
Note that the leading order term coincides with the effective coupling Eq.~(\ref{Effective1+1GluonExchange}) derived from the tree amplitude [in regime (II) or $\Lambda=m_g$ in regime (I)]. This means that the effective coupling at $\Lambda=m_g$ deviates from the tree value and the difference corresponds to the contribution from the quantum fluctuations. Indeed, if we neglect the second term in Eq.~(\ref{RGeq-I}), we will obtain this leading solution.  

If we enter regime (II), we solve the RG equation (\ref{RGeq-II}) with the initial condition specified at $\Lambda=m_g$. In the previous analysis, we solved the RG equation (\ref{RGeq-II}) with the initial condition given by the tree value. However, the initial value must be replaced by the result (\ref{Gmg}) which we have just obtained from the RG equation in regime (I). Therefore, the solution to the RG equation (\ref{RGeq-II}) with a renewed (and more accurate) initial condition reads:
\beq
G(\Lambda)
&=& \frac{ G(m_{g}) }{ 1 + \frac{1}{4\pi} N_{c}  G(m_{g})\, {\rm{log}}( \Lambda / m_{g} )}.
\eeq
Since the functional form is the same as before, we can immediately find that this solution has a Landau pole and we can define the Kondo scale.
By using the $G(m_{g})$ in Eq. (\ref{Gmg}) up to the second order, 
we finally obtain the Kondo scale as
\beq
\Lambda_{\rm K}
&\simeq& \sqrt{e_{q}B} \, \alpha_{s}^{1/2}\, {\rm{exp}} \left\{ - \frac{ {4} \pi }{ N_{c} \alpha_{s} {\rm{log}} \left( {4} \pi / \alpha_{s} \right) } 
+ {\rm{log}} \left( \frac{ { 4 } \pi}{\alpha_{s}} \right)^{{1/6}} \right\}  \nonumber \\
&\simeq& \sqrt{e_{q}B}\, \alpha_{s}^{{ 1/3} }\, {\rm{exp}} \left\{ - \frac{ {4} \pi }{ N_{c} \alpha_{s} {\rm{log}} \left( {4} \pi / \alpha_{s} \right) } \right\}.
\label{Kscale2}
\eeq
The power of $\alpha_{s}$ in the prefactor is now $1/3$ which is slightly smaller than $1/2$ due to the effects of quantum fluctuations from regime (I) $\Lambda > m_g$.
In fact, the exponent of the Kondo scale is predominantly determined by the dynamics in regime (II), while the prefactor by regime (I). This implies that the numerical value of the prefactor depends on the approximations made in regime (I). For example, recall that we have neglected the magnetic phase factor in deriving the RG equation for the effective coupling $G$. While this approximation leads to a closed RG equation for the effective coupling, its effect could become sizable when we consider higher energy region in regime (I). Still, we expect an improved treatment at higher energy region will only affect the prefactor, while keeping the exponent unchanged.

We can show that this Kondo scale (\ref{Kscale2}) is common to several channels of SU$(N_{c})$ symmetry, including $\bar{3}_{c}$ and $6_{c}$ of $N_{c}=3$, as in the QCD Kondo effect \cite{Hattori:2015hka}. Furthermore, we can easily verify that the same Kondo scale appears in the $q$-$\bar{Q}$ scattering channel.

As mentioned in the Introduction, the resultant Kondo scale (\ref{Kscale2}) has the form similar to that of the dynamical quark mass $m_{\rm dyn}$ induced by magnetic catalysis in QCD \cite{Miransky:2002rp}. 
This is not an accidental similarity, and there are at least two reasons. First of all, recall that the analytic representation of the dynamical mass in magnetic catalysis is obtained from the Dyson-Schwinger (DS) equation for a quark. Although the DS equation can in principle contain nonperturbative effect, it was solved in Ref.~\cite{Miransky:2002rp} with an improved ``rainbow" approximation which consists of an infinite number of ``magnetically dressed" gluon propagators attached to a single quark propagator in a planar way. Cutting the (virtual) quark propagator generates planar diagrams for quark-antiquark scattering with infinitely many gluon exchange. Notice that similar types of diagrams can be summed by the RG equation which recursively generates gluon exchange.\footnote{In the Kondo effect, nonplanar crossed diagrams represented in Fig.~\ref{Fig:one-loop} (right) play an important role in the cancelling the spurious contribution from $U(1)$ (scalar) interactions.}
Thus solving the DS equation in the rainbow approximation is intimately related to solving the perturbative RG equation.\footnote{It is recognized in color superconductivity that the gap obtained from the gap equation (the DS equation) in the rainbow approximation is equivalent to that computed from the perturbative RG equation \cite{Hsu:1999mp}.  } 
In both cases, the exponent in the characteristic scale has the scattering amplitude at the tree level.
The secondary reason is the fact that the gluon exchange interaction between the light quark and the heavy quark impurity (the tree level amplitude) in our problem has the form similar to that between the light quark and the anti-light quark in magnetic catalysis. This situation should be contrasted with the relationship between the color superconductivity and the QCD Kondo effect. Although both of them are unstable phenomena around the Fermi surface, the gap in the color superconductivity and the Kondo scale in the QCD Kondo effect have {\it different} parametric dependences on the coupling. This is because the dominant interaction leading to each phenomenon is different from each other: the color {\it magnetic} gluon exchange in the color superconductivity \cite{Son:1998uk} and the color {\it electric} gluon exchange in the QCD Kondo effect \cite{Hattori:2015hka}.

To reach a deeper understanding of the result, let us consider again how the Kondo scale is obtained. What we actually did is the following: We first solve the perturbative RG equation to find the scale dependence of the effective coupling $G(\Lambda)$ in a ``safe" region where the effective coupling $G(\Lambda)$ is small enough. Then we find that $G(\Lambda)$ increases with decreasing $\Lambda$, and diverges at some scale while it stands outside the validity region of perturbative calculation. This scale is nothing but the Kondo scale $\Lambda_{\rm K}$. In this sense, we are able to determine $\Lambda_{\rm K}$ from the behavior of $G(\Lambda)$ even without going down to lower scales. 

On the other hand, we should notice that solving the RG equation corresponds to collecting logarithmically enhanced contributions from each step of the degrading scale. Thus, the effective coupling at scale $\Lambda$ is essentially controlled by the longitudinal integral $\int^{\Lambda_0}_{\Lambda} dK/K$. This means that the physics at the scale $\Lambda$ is determined by contributions from the whole region $\Lambda < K < \Lambda_0$ (instead of the infrared region).

In the expression of the Kondo scale (\ref{Kscale}), we saw that the exponent of the Kondo scale does not have an explicit dependence on the magnetic field strength $B$. The same is true for the improved result (\ref{Kscale2}). If the Kondo scale depends on the magnetic field only through the overall prefactor, then it grows very fast with increasing $B$. However, this is not the case. In fact, a nontrivial dependence appears through the QCD coupling constant $\alpha_s$, which renders the growth slower. 
We take the QCD coupling evaluated at the energy scales of the order of $\sqrt{e_qB}$.
This is consistent with the observation that the effective coupling in the Kondo scale must be evaluated at the initial energy scale $\Lambda_0\sim \sqrt{e_qB}$. Plugging the QCD coupling $\alpha_s(e_{q}B)^{-1} \simeq b_0 \log (e_{q}B/\Lambda_{\rm QCD}^2)$ with $b_0=(11N_c-2n_f)/12\pi$ into the Kondo scale (\ref{Kscale2}), we find the following $B$ dependence: 
\beq
\frac{ \Lambda_{\rm K}^{2}\ \ \, }{ \Lambda_{\rm QCD}^{2} }
&\simeq& 
\frac{ 1 }{\left[ b_{0} {\rm{log}} \left( \frac{ e_{q} B}{ \Lambda_{\rm QCD}^{2} } \right) \right]^{2/3} } 
\left( \frac{ e_{q}B }{ \Lambda_{\rm QCD}^{2} } \right)^{ 
1 - 2\gamma },
\eeq
where the power 
$\gamma\equiv (4\pi b_0/N_c) /
 \log\{4\pi b_0 \log(e_qB/\Lambda_{\rm QCD}^2)\}$ 
corresponds to the anomalous dimension for $\Lambda_{\rm K}(B)$ 
coming from quantum evolution.
This Kondo scale very slowly but monotonically increases with an increasing magnetic field like the dynamical quark mass induced by magnetic catalysis in QCD \cite{Miransky:2002rp}. 
Taking the magnetic field strong enough drives the Kondo scale and thus the physics of the Kondo dynamics far from the nonperturbative QCD scale $\Lambda_{\rm QCD}$. In particular, it is quite interesting that the Kondo effect could take place in the energy region where the QCD coupling is small enough.


\section{Impact of magnetically induced QCD Kondo effect}

In this section, we discuss how the magnetically induced QCD Kondo effect could give an impact on other phenomena that are induced by strong magnetic fields. We also discuss possible relevance of the magnetically induced QCD Kondo effect in realistic situations which are accompanied by strong magnetic fields.

\subsection{Magnetic catalysis vs magnetically-induced QCD Kondo effect}

In the previous analysis for the QCD Kondo effect \cite{Hattori:2015hka}, competition with the color superconductivity was discussed because both are expected to occur in high density quark matter. Light quarks near the Fermi surface induce the Cooper instability via attractive quark-quark interaction in color superconductivity, and the avalanche of particle-hole excitations via quark-impurity (non-Abelian) interaction in the QCD Kondo effect. Thus, how the competition occurs depends on the strength of these two interactions (which may be controlled by the density of impurities). In the current analysis of the magnetically induced QCD Kondo effect, we have taken the chemical potential $\mu$ much larger than the dynamical quark mass $m_{\rm{dyn}}$ so that the effect of finite mass can be neglected above the Fermi level and we can focus on the Kondo dynamics. However, when the chemical potential is taken small and the effect of finite mass cannot be neglected any longer even above the Fermi level, we will have to examine the competition between the magnetically induced QCD Kondo effect and magnetic catalysis. In this case, since the chemical potential and thus the Kondo scale are not far away from $m_{\rm{dyn}}$, the result obtained with the massless approximation should be taken with care. 
Namely when the competition becomes crucial with $\mu$, $m_{\rm dyn}$ and $\Lambda_{\rm K}$ of the similar order, the effective coupling $G$ will become large, and thus we will have to work with some nonperturbative methods.

Then, what kind of physics is expected by the competition? Physically, we expect to see the competition between two effects: one is the interaction to form a chiral condensate (a $q\bar q$ pair) and the other, a Kondo color-singlet state (a $qQ$ pair).
As we already discussed briefly in the Introduction, such a competition will give a sizable influence on the QCD phase diagram with strong magnetic fields. Consider the QCD phase diagram on a temperature ($T$) and magnetic field ($B$) plane. When the magnetic field is not imposed ($B=0$), the chiral symmetry breaking (and quark confinement) takes place in low temperature regions $T<T_c$, but addition of the magnetic field $B\neq 0$ induces the magnetic catalysis and affects the critical temperature in a nontrivial way. For example, the phenomena induced by magnetic fields are enhancement of chiral \cite{Suganuma:1990nn, Gusynin:1994re, Gusynin:1994xp} and gluon \cite{Bali:2013esa, Ozaki:2013sfa} condensates in lower temperature regions (magnetic catalysis), and decrease of $T_c$  of chiral \cite{Bali:2012zg, Bali:2011qj} and deconfinement \cite{Bruckmann:2013oba, Ozaki:2015yja} phase transitions (inverse magnetic catalysis). If one adds heavy quarks to a system that is in chirally broken phase and in magnetic fields, then heavy quarks attract light quarks constituting the chiral condensate, due to the magnetically-induced Kondo effect, and will work to weaken the magnetic catalysis. This  will have some impact on the critical temperature to modify the QCD phase diagram.

\subsection{Chiral magnetic effect in the presence of the heavy impurity}

Recall that the Kondo effect was originally introduced in relation to the anomalous behavior of electric resistance at low temperature. Similarly, we expect that transport properties of the QCD matter will also change due to the QCD Kondo effect. Analogy with the ordinary Kondo effect suggests that the QCD Kondo effect at high densities will affect the electric and color electric conductivities which can be seen as susceptibilities against external fields. Then, what can be expected as a result of the magnetically induced QCD Kondo effect? A naive expectation is that it would affect the electric/color electric conductivities in a similar way as in the QCD Kondo effect at high densities. However, we should recall that while there is a logarithmic enhancement in the scattering amplitude, only a forward scattering is possible with 1+1-dimensional massless quarks. Therefore, the scattering cross section does not contribute to the electrical resistance.\footnote{A standard framework for computing the electrical resistance is the Boltzmann equation with the interaction between an electron and an impurity included in the collision term. In principle, we can do the same thing for the QCD Kondo effect and the magnetically induced QCD Kondo effect.} A similar argument holds in the chiral magnetic effect which is an interesting transport phenomenon in a chirally imbalanced quark matter in a strong magnetic field \cite{Kharzeev:2007jp, Fukushima:2008xe}. Indeed, we already confirmed that the QCD Kondo effect (or a logarithmic enhancement of the scattering amplitude) is also caused by the chiral imbalance \cite{Sho_fu} without the magnetic field. However, if we consider the QCD Kondo effect in the presence of both the chiral imbalance and the strong magnetic fields with the condition $\sqrt{e_{q}B} \gg \mu_{5}$, it is dominated by magnetically induced QCD Kondo effect. 
In this case, only the forward scattering will be allowed due to the dimensional reduction to 1+1-dimensions. Thus, we expect that the chiral magnetic effect will not be affected by this Kondo effect. We will discuss this problem in a separate paper \cite{Sho_fu}.

\subsection{Relevance in realistic situations: 
Heavy ion collisions and magnetars}

There are several situations which are accompanied by extremely strong magnetic fields. The primary example is high-energy heavy-ion collisions at RHIC or the LHC. While the duration for the strong magnetic fields is quite short, it may last as long as the lifetime of quark-gluon plasmas and the typical strength would exceed even the nonperturbative QCD scale $eB \simge \Lambda_{\rm QCD}^2$. The secondary example is the magnetars whose magnetic fields well exceed the critical magnetic field for electrons $eB\gg eB_c=m_e^2$. We expect that a quark matter could exist at the core of magnetars since the density goes beyond the normal nuclear matter density. In both cases, if heavy quark impurities are present, there is a possibility for the magnetically induced Kondo effect to occur. As we already discussed, the appearance of the Kondo effect will be seen in the change of transport properties of the quark matter.

\subsection{Applications to condensed matter physics}

In certain systems in condensed matter physics, 
the electronic spectrum takes the form of chiral fermions.
For example, carbon nanotube has a 1+1-dimensional spectrum, and graphene has a 2+1-dimensional spectrum. 
The parent crystal is long-known as graphite that is composed of infinite stacking of graphene layers.  
Because of weak interlayer interaction, the electronic spectrum of graphite has a dispersion also along the c-axis that is perpendicular to the layers, and the Fermi surface of electrons and holes is present.  
If one applies a strong magnetic field parallel to the c-axis, 
the carriers fall into the LLL, and their spin is polarized. 
However, there remains still a sublattice degrees of freedom that plays a role analogous to the color of quarks.
Hence, interesting many-body phenomena can be expected.  
In fact, recent experiment under magnetic field of the order of 50 T has found a feature suggesting  drastic change of the electronic property \cite{tokunaga:2015}.
With impurity scatterings from one sublattice to the other, we may also expect a Kondo-type effect that is analogous to the magnetically induced QCD Kondo effect discussed in this paper.


\section{summary and conclusion}

We have found a new type of the Kondo effect, the ``magnetically induced QCD Kondo effect" which occurs when a strong magnetic field is imposed on a light quark matter with heavy-quark impurities. In this Kondo effect, imposing the magnetic field plays the role as forming the Fermi surface. We understand that the magnetically induced QCD Kondo effect shares the same fundamental physics with superconductivity and magnetic catalysis. We have explicitly demonstrated that, in a strong magnetic field, the scattering amplitude of a massless quark off a heavy quark impurity shows a logarithmic enhancement with respect to decreasing energy scale. The effective coupling strength between a light quark and a heavy impurity shows a typical behavior of the asymptotic freedom, and thus diverges at a lower scale called the Kondo scale. It is given by  $\Lambda_{\rm K} \simeq \sqrt{e_qB}\  \alpha_{s}^{1/3} {\rm{exp}}\{- {4} \pi/N_{c} \alpha_{s} {\rm{log}}({4} \pi/\alpha_{s}) \}$ which has the form similar to the dynamical mass of magnetic catalysis in QCD \cite{Miransky:2002rp}. The Kondo scale slowly grows with increasing magnetic field strength and can become larger than the QCD nonperturbative scale $\Lambda_{\rm QCD}$. Therefore, a sufficiently strong magnetic field allows us to find the Kondo effect in a regime where the QCD coupling is small enough. Since only the forward scattering is allowed due to the dimensional reduction in a strong magnetic field, the magnetically induced QCD Kondo effect will not affect the electrical resistance. However, its effect can be seen in other transport properties.

\section*{Acknowledgments}
We thank S.~Yasui for helpful discussions and valuable comments. 
S. O. is also grateful to I.A.~Shovkovy for useful discussion and important comments.
This work was supported in part by the Center for the Promotion of Integrated Sciences (CPIS) of Sokendai.

\appendix

\section{Calculation of dimensionally reduced scattering amplitudes}

\begin{figure}
\begin{minipage}{0.9\hsize}
\begin{center}
\includegraphics[width=0.7 \textwidth]{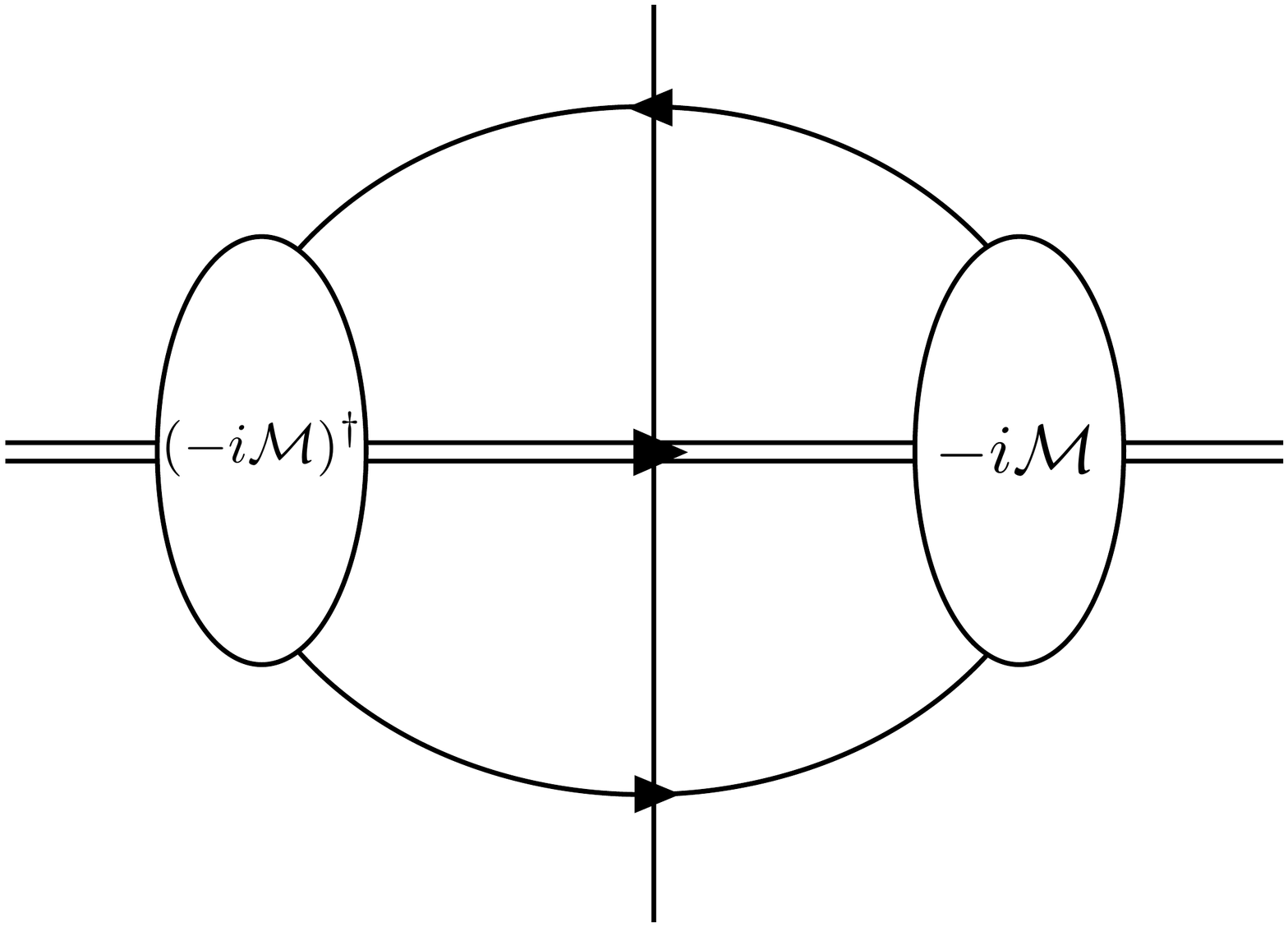}
\vskip -0.1in
\end{center}
\end{minipage}
\caption{
Self-energy diagram of the heavy quark.
}\label{Fig:selfenergy}
\end{figure}

\begin{figure}
\begin{minipage}{0.8\hsize}
\begin{center}
\includegraphics[width=0.7 \textwidth]{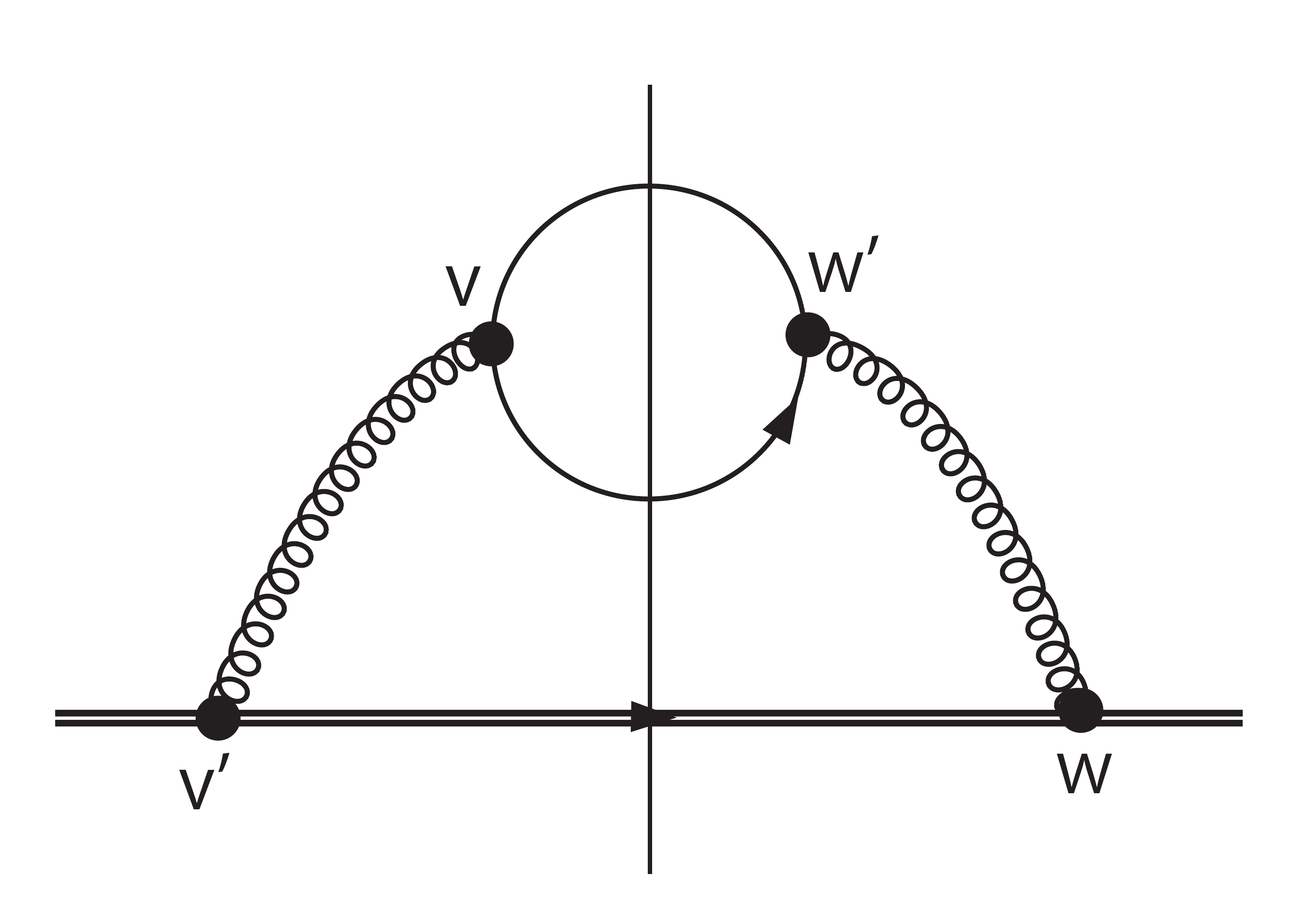}
\vskip -0.1in
\end{center}
\end{minipage}
\caption{
Leading order contribution to Fig.~4.
}\label{Fig:LO}
\end{figure}

In the Appendix, we present how to obtain the dimensionally reduced amplitude for the scattering of a light quark in the LLL state off a heavy quark impurity. We perform perturbative calculation with respect to the QCD coupling because we have large scales $\sqrt{e_{q}B} \gg \mu \gg \Lambda_{\rm QCD}$ which provide typical energy scales of the coupling $\alpha_s(\sqrt{e_qB})\ll 1$. Below, we compute the tree (leading order) and one-loop (next-to-leading-order) amplitudes through the self-energy of a heavy quark, which is a technique adopted also in Ref.~\cite{Fukushima:2015wck}. The self-energy of a heavy quark with a one-loop contribution of a light quark as depicted in Fig.~\ref{Fig:selfenergy} is directly related to the amplitude of the scattering of a light quark off the heavy quark. In other words, the self-energy $\Sigma$ is proportional to $|\mathcal{M}|^2$ and the amplitude $\mathcal{M}$ can be read off through the cutting rule. 
Our method in a strong magnetic field can derive the gauge invariant result with minimum complication of phase factors due to the magnetic field.

To demonstrate how to extract the scattering amplitude from the self-energy diagram, we first consider the case of vanishing magnetic field.
In this case, the leading order diagram is shown in Fig.~\ref{Fig:LO} and the self-energy $\Sigma^{\rm LO}(r,r'|B=0,\mu\neq 0)$ in the coordinate space is given by
\beq
&&\hspace{-1cm}\Sigma^{\rm LO}(0,0|B=0,\mu\neq 0)\nonumber\\
&\equiv&
\int d^{4}w d^{4}w^{\prime} d^{4}v d^{4}v^{\prime}
\ {\rm{tr}} \left( S(v,w^{\prime}) ig \gamma^{\mu} (T^{A})_{a^{\prime} a} S(w^{\prime}, v) ig \gamma^{\alpha} (T^{A^{\prime}})_{a a^{\prime}} \right) \nonumber \\
&& \hspace{-3mm}\times D^{AB}_{\mu \nu}(w, w^{\prime}) D^{A^{\prime} B^{\prime}}_{\alpha \beta}(v, v^{\prime})
S_{\rm H}(0, w) ig \gamma^{\nu} (T^{B})_{b^{\prime}b }S_{\rm H} ( w, v^{\prime}) ig \gamma^{\beta} (T^{B^{\prime}})_{b c} S_{\rm H}(v^{\prime}, 0) \nonumber \\
&=& \int_{q^{\prime}, q } \ {\rm{tr}} \left( \tilde{S}(q^{\prime};\mu) ig \gamma^{\mu} (T^{A})_{a^{\prime} a} \tilde{S}(q;\mu) ig \gamma^{\alpha} (T^{A^{\prime}})_{a a^{\prime}} \right) \nonumber \\
&& \hspace{-3mm}\times \tilde D^{AB}_{\mu \nu} (q^{\prime}-q;\mu) \tilde D^{A^{\prime} B^{\prime} }_{\alpha \beta} ( q^{\prime} - q;\mu )
\int_{P, P^{\prime}} \!\!\! \tilde{S}_{\rm H}(P^{\prime}) ig \gamma^{\nu} (T^{B})_{b^{\prime} b} \tilde{S}_{\rm H}(P) ig \gamma^{\beta} (T^{B^{\prime}})_{b c} \tilde{S}_{\rm H}(P^{\prime}), 
\label{4_dim_tree}
\eeq
where $S(r,r')$, $S_{\rm{H}}(r, r^{\prime})$ and $D_{\mu\nu}^{AB}(r,r')$ are in-medium propagators in coordinate space of a light quark, a heavy quark and a gluon, and we put tilde for their counterparts in the momentum space. 
The momentum integrals in the rightmost expression stand for the four momentum ones: $\int_q=\int d^4q/(2\pi)^4$. 
For simplicity, we have set the two coordinates of the external heavy quarks equal to zero. By applying the cutting rule to the intermediate quark states, and replacing the external heavy propagator by the spinor, we find the following amplitude:
\beq
-i \mathcal{M}_{0}(B=0, \mu\neq 0)\nonumber\\
&&\hspace{-3cm}= (ig)^{2}\left[\bar{u}(q^{\prime}) \gamma^{\mu}(T^{A})_{a^{\prime} a} u(q)\right] \tilde D^{AB}_{\mu \nu}(q^{\prime}-q;\mu) \left[\bar{U}(P^{\prime}) \gamma^{\nu} (T^{B})_{b^{\prime}b }U(P)\right].
\label{scattering_at_zeroB}
\eeq
This is nothing but the leading-order scattering amplitude of the light quark off the heavy quark at a finite chemical potential and vanishing magnetic fields.
This leading order amplitude is the same as that in the analysis of the QCD Kondo effect \cite{Hattori:2015hka}. In the absence of the magnetic field, this amplitude contains full dependence on the four momenta of the scattered quark.\\

\noindent {\bf 1. Leading order}

Now, we shall compute the scattering amplitudes in strong magnetic fields within the leading-order level. 
In this case, the light quarks are in the LLL state, and the gluons are screened due to the magnetic field (recall $\sqrt{e_{q}B} \gg \mu$). Then, the leading-order self-energy of a heavy quark shown in Fig.~5 can be written as
\beq
\Sigma^{\rm LO}(0,0|B,\mu)\nonumber\\
&&
\hspace{-2cm}=\int d^{4}w d^{4}w^{\prime} d^{4}v d^{4}v^{\prime}
\ {\rm{tr}} \left( \mathcal{S}_{\rm{LLL}}(v ,w^{\prime}) ig\gamma^{\mu} (T^{A})_{a^{\prime}a} \mathcal{S}_{\rm{LLL}} (w^{\prime}, v) ig \gamma^{\alpha} (T^{B^{\prime}})_{a a^{\prime}} \right) \nonumber \\
&&\hspace{-2cm} \times \mathcal{D}^{AB}_{\mu \nu} (w, w^{\prime}|e_{q}B) \mathcal{D}^{A^{\prime} B^{\prime}}_{\alpha \beta} (v, v^{\prime}|e_{q}B)
S_{\rm H}(0, w) ig \gamma^{\nu} (T^{B})_{b^{\prime} b} S_{\rm H} ( w, v^{\prime}) ig \gamma^{\beta} (T^{B^{\prime}})_{b c} S_{\rm H}(v^{\prime}, 0) \nonumber \\
&&\hspace{-2cm}= \frac{ e_{q}B}{ 2\pi } \int_{q_{\parallel}^{\prime}, q_{\parallel} } \ {\rm{tr}} \left( \tilde{S}_{\parallel}(q^{\prime}_{\parallel};\mu) ig \gamma^{\mu} (T^{A})_{a^{\prime} a} \tilde{S}_{\parallel} (q_{\parallel};\mu) ig \gamma^{\alpha} (T^{A^{\prime}})_{a a^{\prime}} \right) \nonumber \\
&&\hspace{-2cm} \times \int_{Q_{\perp}} \tilde{\mathcal{D}}^{AB}_{\mu \nu } ( q_{\parallel}^{\prime} - q_{\parallel}, Q_{\perp}|e_{q}B)\, {\rm{e}}^{-\frac{l_{q}^{2} Q_{\perp}^{2} }{ 4 } }
\int_{\bar{Q}_{\perp}} \tilde{\mathcal{D}}^{A^{\prime} B^{\prime} }_{\alpha \beta } ( q_{\parallel}^{\prime} - q_{\parallel}, \bar{Q}_{\perp}|e_{q}B)\, {\rm{e}}^{-\frac{l_{q}^{2} \bar{Q}_{\perp}^{2} }{ 4 } } \nonumber \\
&&\hspace{-2cm} \times \int_{P, P^{\prime}} \tilde{S}_{\rm H}(P^{\prime}) ig \gamma^{\nu} (T^{B})_{b^{\prime} b} \tilde{S}_{\rm H}(P) ig \gamma^{\beta} (T^{B^{\prime}})_{bc} \tilde{S}_{\rm H}(P^{\prime}),
\label{self_leading}
\eeq
where the momentum integrals are defined as $\int_{q_{\parallel}^{(\prime)} } = \int \frac{ d^{2} q_{\parallel}^{(\prime)} }{ (2\pi)^{2} }$, $\int_{Q_{\perp}} = \int \frac{ d^{2} Q_{\perp} }{ (2\pi)^{2} }$,
and $\int_{P^{(\prime)}} = \int \frac{ d^{4} P^{(\prime)} }{ (2\pi)^{4} }$.
Comparing with the previous simple case without magnetic fields (\ref{4_dim_tree}) where the four coordinates $v,v',w,w'$ are replaced by four momenta $q,q',P,P'$, the transverse momenta of the light quark $q_\perp,q_\perp'$ in the above result are already integrated out leaving only the $Q_\perp, Q_\perp'$ integrals left. Recall that the transverse motion of a light quark in a strong magnetic field can be factored out in the self-energy yielding an overall factor $\frac{ e_{q}B }{ 2\pi }$. This factorization property corresponds to the dimensional reduction in a strong magnetic field. Below, we exclude this overall factor in defining the dimensionally reduced amplitude.
By using the cutting rule as in the previous simple example, and replacing the external heavy quark propagator by the spinor, we can read off the leading order (1+1 dimensional) scattering amplitude as 
\beq
-i \mathcal{M}^{\rm{LLL}}_{0}\nonumber\\
&&\hspace{-15mm}=
(ig)^{2} \left[\bar{u}_{\rm LLL}(q^{\prime}_{\parallel}) \gamma^{\mu} (T^{A})_{a^{\prime} a} u_{\rm LLL}(q_{\parallel})\right]
\int_{Q_{\perp}} \tilde{\mathcal{D}}^{AB}_{\mu \nu } ( q_{\parallel}^{\prime} - q_{\parallel}, Q_{\perp}|e_{q}B) {\rm{e}}^{- \frac{ l_{q}^{2} Q_{\perp}^{2} }{ 4 } } 
\left[\bar{U}(P^{\prime}) \gamma^{\nu} (T^{B})_{b^{\prime}b} U(P)\right]. \nonumber \\
\eeq
This is the tree amplitude (\ref{tree}). \\

\begin{figure}
\begin{minipage}{0.8\hsize}
\begin{center}
\includegraphics[width=0.7 \textwidth]{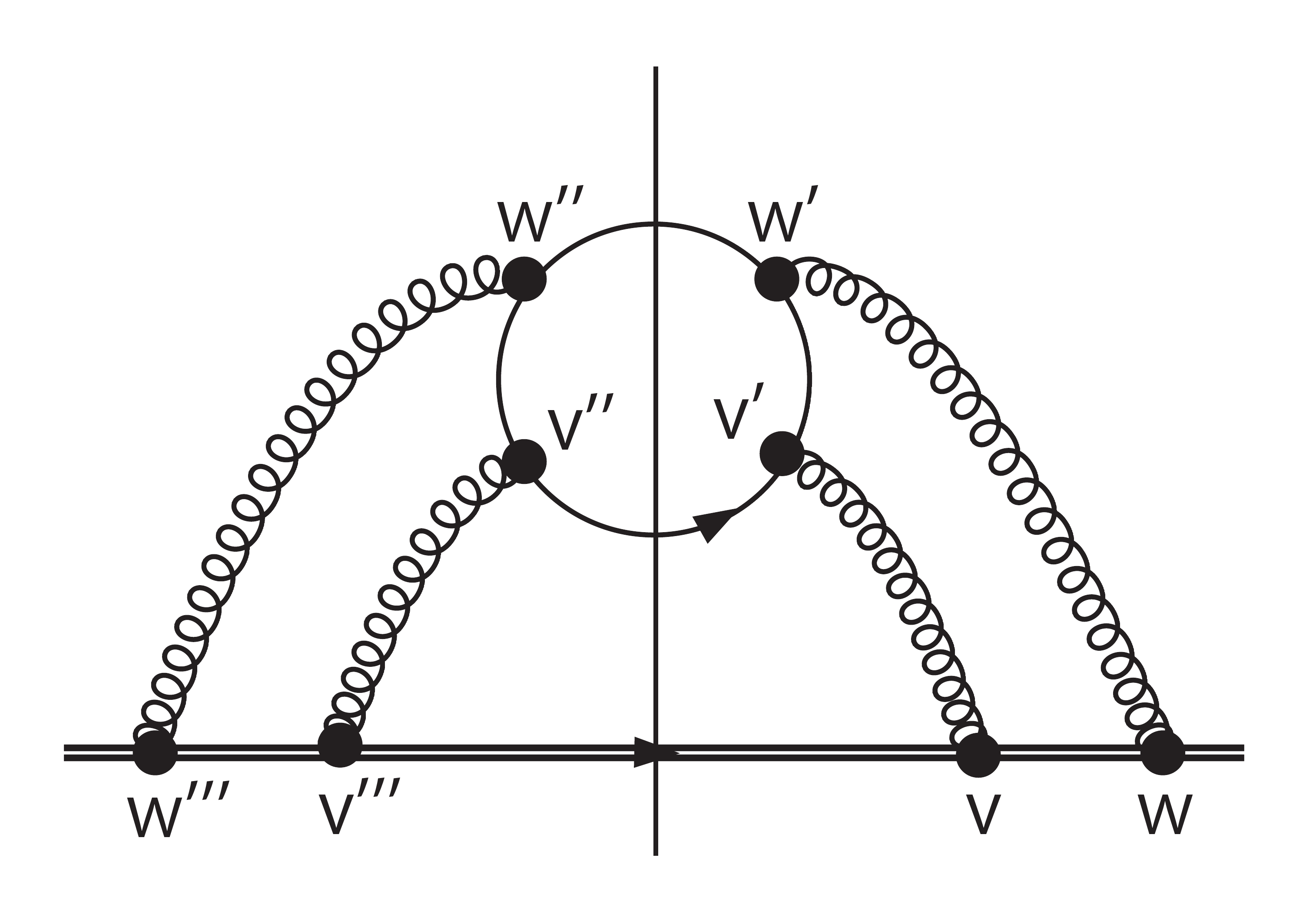}
\vskip -0.1in
\end{center}
\end{minipage}
\caption{
Next to leading order contribution to Fig.~\ref{Fig:selfenergy}, corresponding to the box diagram.
}\label{Fig:NLO}
\end{figure}

\noindent {\bf 2. Next to leading order}

Next, we consider the next-to-leading order processes which consist of box and crossed diagrams. The self-energy diagram leading to the box diagram is depicted in Fig.~\ref{Fig:NLO}, and the self-energy $\Sigma^{\rm NLO(box)}(r, r'|B,\mu)$ with two coordinates taken to be zero $r=r'=0$ is given by 
\beq
\Sigma^{\rm NLO(box)}(0,0|B,\mu)\nonumber\\
&&
\hspace{-40mm}=\int d^{4}w d^{4}w^{\prime} d^{4}w^{\prime \prime} d^{4} w^{\prime \prime \prime}
d^{4}v d^{4}v^{\prime} d^{4} v^{\prime \prime} d^{4} v^{\prime \prime \prime} \nonumber \\
&&\hspace{-35mm} \times {\rm{tr}} \Big( \mathcal{S}_{\rm{LLL}} ( w^{\prime \prime}, w^{\prime} ) ig \gamma^{\mu} (T^{A})_{a^{\prime} a^{\prime \prime}} \mathcal{S}_{\rm{LLL}} (w^{\prime}, v^{\prime} ) ig \gamma^{\nu} (T^{B})_{a^{\prime \prime} a}
\nonumber\\
&&\left. \times 
 \mathcal{S}_{\rm{LLL}} ( v^{\prime}, v^{\prime \prime} ) ig \gamma^{\xi} (T^{A^{\prime}})_{a \bar{a}} \mathcal{S}_{\rm{LLL}} (v^{\prime \prime}, w^{\prime \prime} ) ig \gamma^{\alpha} (T^{B^{\prime}})_{\bar{a} a^{\prime}} \right) \nonumber \\
&&\hspace{-35mm} \times \mathcal{D}^{AC}_{\mu \sigma} (w, w^{\prime}) \mathcal{D}^{BD}_{\nu \rho} ( v, v^{\prime} ) \mathcal{D}^{A^{\prime} C^{\prime} }_{\xi \delta} ( v^{\prime \prime}, v^{\prime \prime \prime} ) \mathcal{D}^{B^{\prime} D^{\prime}}_{\alpha, \beta} (w^{\prime \prime}, w^{\prime \prime \prime} ) \nonumber \\
&&\hspace{-35mm} \times S_{\rm{H}}(0, w) ig \gamma^{\sigma} (T^{C})_{b^{\prime} b^{\prime \prime} } S_{\rm{H}} (w, v ) ig \gamma^{\rho} (T^{D})_{b^{\prime \prime} b} 
\nonumber\\
&&
\times S_{\rm{H}} ( v, v^{\prime \prime \prime} ) ig \gamma^{\delta} (T^{C^{\prime}})_{b c} S_{\rm{H}} (v^{\prime \prime \prime}, w^{\prime \prime \prime}) ig\gamma^{\beta} (T^{D^{\prime}})_{cd} S_{\rm{H}} ( w^{\prime \prime \prime} ,0 ) \nonumber \\
&&\hspace{-40mm}= \frac{e_{q}B}{2\pi} \int_{q^{\prime}_{\parallel}, q_{\parallel} } \int_{k_{\parallel}, \bar{k}_{\parallel}} \nonumber\\
&&\hspace{-35mm}\times {\rm{tr}} \left(  \tilde{S}_{\parallel} (q^{\prime}_{\parallel};\mu ) ig \gamma^{\mu} (T^{A})_{a^{\prime} a^{\prime \prime} } \tilde{S}_{\parallel} (k_{\parallel};\mu) ig \gamma^{\sigma} (T^{B})_{a^{\prime \prime}a} \tilde{S}_{\parallel} ( q_{\parallel} ; \mu) ig \gamma^{\xi} (T^{A^{\prime}})_{a \bar{a}} \tilde{S}_{\parallel} (\bar{k}_{\parallel};\mu) ig \gamma^{\alpha} (T^{B^{\prime}})_{\bar{a} a}
\right) \nonumber \\
&&\hspace{-35mm} \times \left[ \int_{Q_{\perp}^{\prime} } \tilde{\mathcal{D}}^{AC}_{\mu \sigma} (q^{\prime}_{\parallel} - k_{\parallel}, Q_{\perp}^{\prime} |e_{q}B) {\rm{e}}^{ - \frac{ l_{q}^{2} Q_{\perp}^{\prime \ 2} }{ 4 } } 
 \int_{Q_{\perp}} \tilde{\mathcal{D}}^{BD}_{\nu \rho} (q_{\parallel} - k_{\parallel}, Q_{\perp} |e_{q}B) {\rm{e}}^{ - \frac{ l_{q}^{2} Q_{\perp}^{2} }{ 4 } } \right]
{\rm{e}}^{ - i \frac{ l_{q}^{2} \vec{Q}_{\perp} \times \vec{Q}_{\perp}^{\prime} }{ 2 } } \nonumber \\
&&\hspace{-35mm} \times \left[ 
\int_{\bar{Q}_{\perp}^{\prime} } \tilde{\mathcal{D}}^{A^{\prime} C^{\prime}}_{\xi \delta} ( q_{\parallel}- \bar{k}_{\parallel}, \bar{Q}_{\perp}^{\prime} | e_{q} B ) {\rm{e}}^{ - \frac{ l_{q}^{2} \vec{ \bar{Q} }_{\perp}^{\prime \ 2} }{ 4} }
\int_{\bar{Q}_{\perp} } \tilde{\mathcal{D}}^{B^{\prime} D^{\prime}}_{\alpha \beta} ( q^{\prime}_{\parallel} - \bar{k}_{\parallel}, \bar{Q}_{\perp} | e_{q} B ) {\rm{e}}^{ - \frac{ l_{q}^{2} \vec{ \bar{Q} }_{\perp}^{2} }{ 4} }
\right] {\rm{e}}^{  i \frac{ l_{q}^{2} \vec{\bar{Q}}_{\perp} \times \vec{\bar{Q}}_{\perp}^{\prime} }{ 2 } } \nonumber \\
&&\hspace{-35mm} \times \int_{P, P^{\prime}} \tilde{S}_{\rm{H}} (P^{\prime}) ig \gamma^{\nu} (T^{C})_{b^{\prime} b^{\prime \prime} } \tilde{S}_{\rm{H}} ( P + q - k ) ig \gamma^{\rho} (T^{D})_{b^{\prime \prime} b} \nonumber\\
&&\times  \tilde{S}_{\rm{H}} ( P )
\gamma^{\delta} (T^{C^{\prime}})_{bc} \tilde{S}_{\rm{H}} ( P + q - \bar{k} ) ig \gamma^{\beta} (T^{D^{\prime}})_{cd} \tilde{S}_{\rm{H}} ( P^{\prime} )\, . 
\eeq 
Again, the factor $\frac{ e_{q}B }{ 2\pi }$ appears owing to the perpendicular part of the light quark propagators, and we exclude it when defining the 1+1 dimensional scattering amplitude. 
Then we obtain the one-loop amplitude of the box diagram by cutting the diagram, which reads 
\beq
-i \mathcal{M}^{\rm{(a)LLL}}_{\rm{1-loop}}\nonumber\\
&&\hspace{-2cm}=
(ig)^{4} \int_{k_{\parallel}} \left[\bar{u}_{\rm{LLL}} (q^{\prime}_{\parallel}) \gamma^{\mu} (T^{A})_{a^{\prime} a^{\prime \prime} } \tilde{S}_{\parallel} (k_{\parallel};\mu) \gamma^{\nu}
(T^{B})_{a^{\prime \prime} a}  u_{\rm{LLL}}(q_{\parallel}) \right]\nonumber \\
&&\hspace{-2cm}\times  \int_{Q_{\perp}^{\prime} } \tilde{\mathcal{D}}^{AC}_{\mu \sigma} (q^{\prime}_{\parallel} - k_{\parallel}, Q_{\perp}^{\prime} |e_{q}B) {\rm{e}}^{ - \frac{ l_{q}^{2} Q_{\perp}^{\prime \ 2} }{ 4 } } 
\int_{Q_{\perp}} \tilde{\mathcal{D}}^{BD}_{\nu \rho} (q_{\parallel} - k_{\parallel}, Q_{\perp} |e_{q}B) {\rm{e}}^{ - \frac{ l_{q}^{2} Q_{\perp}^{2} }{ 4 } } {\rm{e}}^{ - i \frac{ l_{q}^{2} \vec{Q}_{\perp} \times \vec{Q}_{\perp}^{\prime} }{ 2 } } \nonumber \\
&&\hspace{-2cm} \times \left[\bar{U}(P^{\prime}) \gamma^{\sigma}(T^{C})_{b^{\prime} b^{\prime \prime }}  \tilde{S}_{\rm{H}} ( P + q -k ) \gamma^{\rho} (T^{D})_{b^{\prime \prime} b} U(P)\right]\, .
\eeq
This is the result (\ref{one-loop-amp-a}). 
The phase factor ${\rm{e}}^{ - i \frac{ l_{q}^{2} \vec{Q}_{\perp} \times \vec{Q}_{\perp}^{\prime} }{2} }$ is gauge invariant and in general contributes to the scattering amplitude. 
Similarly, we can obtain the 1+1 dimensional one-loop amplitude of the crossed diagram. The result is given in Eq. (\ref{one-loop_amp-b}).

\end{document}